\documentclass[12pt]{article}

\author{Yu.~M.~Zinoviev
       \thanks{E-mail address: Yurii.Zinoviev@ihep.ru} \\
        {\it Institute for High Energy Physics} \\
        {\it Protvino, Moscow Region, 142280, Russia}}
\title{Towards frame-like gauge invariant formulation\\
for massive mixed symmetry bosonic fields. II.\\
General Young tableau with two rows}

\date{}

\begin{document}

\maketitle

\begin{abstract}
In this paper we complete our construction of frame-like gauge
invariant description for massive mixed symmetry tensor fields
corresponding to arbitrary Young tableau with two rows started in
\cite{Zin08c}. We consider general massive theory in $(A)dS$ spaces
with arbitrary cosmological constant as well as all special limits
which exist both in de Sitter and in anti-de Sitter spaces.
\end{abstract}

\thispagestyle{empty}
\newpage
\setcounter{page}{1}

\setlength{\unitlength}{1mm}

\section*{Introduction}

In this paper we complete our construction of frame-like gauge
invariant description for massive mixed symmetry tensor fields
corresponding to arbitrary Young tableau with two rows started in
\cite{Zin08c}. Frame-like formalism \cite{Vas80,LV88,Vas88} is a
natural generalization of well-known frame formulation of gravity in
terms of veilbein $e_\mu{}^a$ and Lorentz connection
$\omega_\mu{}^{ab}$ and it turns out to be very convenient for the
investigation of possible interacting theories for higher spin
particles as well as of gauge symmetry algebras behind them.

There are two different frame-like formulations for massless mixed
symmetry bosonic fields. For simplicity, let us restrict ourselves
with mixed symmetry tensors corresponding to Young tableau with two
rows. Let us denote $Y(k,l)$ a tensor $\Phi^{a_1 \dots a_k,b_1 \dots
b_l}$ which is symmetric both on first $k$ as well as last $l$
indices, completely traceless on all indices and satisfies a
constraint $\Phi^{(a_1 \dots a_k,b_1) b_2 \dots b_l} = 0$, where round
brackets mean symmetrization. In a metric-like formulation such field
has two gauge transformations with gauge parameters corresponding to
$Y(k,l-1)$ and $Y(k-1,l)$ (the only exception being the case $k=l$
with one gauge symmetry with parameter $Y(k,k-1)$ only).  In the first
approach \cite{ASV03,Alk03,ASV05,ASV06} for the description of
$Y(k,l)$ tensor ($k \ne l$) one use a one-form $e_\mu{}^{Y(k-1,l)}$ as
a main physical field. In this, only one of two gauge symmetries is
realized explicitly and such approach is very well adapted for the
$AdS$ spaces. Another formulation \cite{Skv08} uses two-form
$e_{\mu\nu}{}^{Y(k-1,l-1)}$ as a main physical field in this, both
gauge symmetries are realized explicitly. Such formalism works in flat
Minkowski space while deformation into $AdS$ space requires
introduction of additional fields \cite{BMV00}. Technical reason is
that it is impossible to deform into $(A)dS$ space keeping both gauge
symmetries, while physical reason is that massless fields in $(A)dS$
space in general have more physical degrees of freedom than in flat
Minkowski space. As a byproduct of our investigation in Sections 2 and
3 we obtain appropriate sets of such additional physical fields both
for $AdS$ and $dS$ spaces.

For completeness we give in Section 1 all necessary information on
massless fields that we will need in the following sections for
constructing massive theories. Also we consider in this section a
possibility to deform such massless theories into $(A)dS$ space. In
most cases such deformation is impossible but the structure of
mass-like terms and corresponding corrections for gauge
transformations will be heavily used in what follows.

Then we turn to the construction of gauge invariant formulation for
massive fields. Recall that there are two general approaches to
gauge invariant description of massive fields. One of them uses
powerful BRST approach 
\cite{BK05,BKRT06,BKL06,BKR07,BKT07,MR07,BKR08,BKR09}.
Another one, which we will follow in this work,
\cite{KZ97,Zin01,Med03,Met06,Zin08b,Zin08c,BG08,Zin09a,Zin09b}
is a generalization to higher spin fields of well-known mechanism of
spontaneous gauge symmetry breaking. In this, one starts with
appropriate set of massless fields with all their gauge symmetries and
obtain gauge invariant description for massive field as a smooth
deformation. One of the nice features of gauge invariant formulation
for massive fields is that it allows us effectively use all known
properties of massless fields serving as building blocks.
As we have already seen in all cases considered previously and we will
see again in this paper, gauge invariant description of massive fields
always admits smooth deformation into $(A)dS$ space without
introduction of any additional fields besides those that are necessary
in flat Minkowski space so that restriction mentioned above will not
be essential for us.

In Section 2 we consider special case of $Y(k+1,k+1)$ tensor fields
corresponding to rectangular Young tableau and then in Section 3 we
consider general case $Y(k+1,l+1)$ with $k \ne l$. As we will see in
both cases gauge invariance completely fixes all parameters in the
Lagrangian and gauge transformations leaving us with one free
parameter having dimension of mass only. It is hardly possible to give
meaningful definition of what is mass for mixed symmetry (spin)-tensor
fields in $(A)dS$ spaces and we will not insist on any such
definition. Instead, we will simply use this parameter to analyze all
possible special limits that exist in $(A)dS$ spaces. In this, only
fields having the same number of degrees of freedom as massless one in
flat Minkowski space we will call massless ones, while all other
special limits that appear in $(A)dS$ spaces will be called partially
massless \cite{DW01,DW01a,DW01c,Zin01,SV06}.

\section{Massless fields}

In this section we provide all necessary information on massless
fields that we will need in the following sections for constructing of
massive ones. In all cases we consider a possibility to deform
massless theory into $(A)dS$ space. As is well known \cite{BMV00}, in
general such deformation is impossible without introduction of
additional fields, but the structure of mass-like terms and
corresponding corrections for gauge transformations will be heavily
used in the following sections. Note that all Lagrangians in this and
the following sections will be completely antisymmetric on world
indices, thus working in $(A)dS$ spaces we will need so-called Lorentz
covariant derivatives acting on local indices only. Our conventions on
such covariant derivatives will be:
$$
[ D_\mu, D_\nu ] \xi^a = - \kappa (e_\mu{}^a \xi_\nu - e_\nu{}^a
\xi_\mu), \qquad \kappa = \frac{2 \Lambda}{(d-1)(d-2)}
$$

\subsection{Tensor $Y(k+1,0)$} 

Besides mixed symmetry tensor fields we will need frame-like
formulation for completely symmetric tensor fields \cite{Vas80,LV88}
(see also \cite{Zin08b}). The main physical field is a one form
$\Phi_\mu{}^{a_1 \dots a_k} = \Phi_\mu{}^{(a_k)}$ (here and in what
follows we will use the same condensed notations for tensor objects as
in our previous works on the subject \cite{Zin08b,Zin08c,Zin09b}),
completely symmetric and traceless on local indices and auxiliary one
form $\omega_\mu{}^{a,(a_k)}$, symmetric on last $k$ indices,
completely traceless on all local indices and satisfying a constraint
$\omega_\mu{}^{(a,a_k)} = 0$. The Lagrangian and gauge transformations
for massless field have the form:
\begin{eqnarray}
(-1)^k {\cal L}_0 &=& -
\left\{ \phantom{|}^{\mu\nu}_{ab} \right\}
 [ \omega_\mu{}^{e,a(a_{k-1})} \omega_\nu{}^{e,b(a_{k-1})} + 
\frac{1}{k} \omega_\mu{}^{a,(a_k)} \omega_\nu{}^{b,(a_k)} ] +
\nonumber \\
 && + 2  \left\{ \phantom{|}^{\mu\nu\alpha}_{abc} \right\}
\omega_\mu{}^{a,b(a_{k-1})} \partial_\nu \Phi_\alpha{}^{c(a_{k-1})}
\end{eqnarray}
where $\left\{ \phantom{|}^{\mu\nu}_{ab} \right\} = e^\mu{}_a
e^\nu{}_b - e^\mu{}_b e^\nu{}_a$ and so on,
\begin{equation}
\delta_0 \Phi_\mu{}^{(a_k)} = \partial_\mu \zeta^{(a_k)} +
\chi_\mu{}^{(a_k)}, \qquad \delta_0 \omega_\mu{}^{a,(a_k)} =
\partial_\mu  \chi^{a,(a_k)}
\end{equation}
where parameters $\zeta$ and $\chi$ have the same properties on local
indices as $\Phi$ and $\omega$, correspondingly.

Now let us consider deformation into $AdS$ space. If we replace all
derivatives in the Lagrangian and gauge transformations by the
covariant ones, the Lagrangian cease to be invariant:
$$
\delta_0 {\cal L}_0 = (-1)^k \frac{2(k+1)(d+k-3)}{k} \kappa
[ \omega_\mu{}^{\mu,(a_k)} \zeta^{(a_k)}
 - \chi^{\mu,(a_k)} \Phi_\mu{}^{(a_k)} ]
$$
To compensate for this non-invariance we introduce mass-like terms
into the Lagrangian:
\begin{equation}
(-1)^k \Delta {\cal L}_0 = b_k 
\left\{ \phantom{|}^{\mu\nu}_{ab} \right\}
\Phi_\mu{}^{a(a_{k-1})} \Phi_\nu{}^{b(a_{k-1})}
\end{equation}
as well as corresponding corrections for gauge transformations:
\begin{eqnarray}
\delta_2 \omega_\mu{}^{a,(a_k)} &=& \frac{b_k}{(k+1)(d-2)}
[ k e_\mu{}^a \zeta^{(a_k)} - e_\mu{}^{(a_1} \zeta^{a_{k-1})a} +
\nonumber \\
 && + \frac{1}{(d+k-3)} ( 2 g^{(a_1a_2} \zeta_\mu{}^{a_{k-2})a}
- (k-1) g^{a(a_1} \zeta_\mu{}^{a_{k-1})} )  ]
\end{eqnarray}
where $b_k = (k+1)(d+k-3) \kappa$.

\subsection{Tensor $Y(k+1,1)$}

This tensor turns out to be special and requires separate
consideration \cite{Skv08,Zin08c}. Main physical field now --- two
form $\Psi_{\mu\nu}{}^{(a_k)}$ completely symmetric and traceless on
all local indices, while auxiliary field is a one form
$\Omega_\mu{}^{(a_k),bc}$ symmetric on first $k$ indices,
antisymmetric on last two ones, traceless on all local indices and
satisfying a constraint $\Omega_\mu{}^{(a_k,b)c} = 0$. Lagrangian and
gauge transformations for massless field have the form:
\begin{eqnarray}
(-1)^k {\cal L}_0 &=&   
\left\{ \phantom{|}^{\mu\nu}_{ab} \right\} [
\Omega_\mu{}^{a(a_{k-1}),cd} \Omega_\nu{}^{b(a_{k-1}),cd} + 
\frac{2}{k} \Omega_\mu{}^{(a_k),ac} \Omega_\nu{}^{(a_k),bc} ] -
   \nonumber \\
 && - \left\{ \phantom{|}^{\mu\nu\alpha\beta}_{abcd} \right\}
\Omega_\mu{}^{a(a_{k-1}),bc} \partial_\nu
\Psi_{\alpha\beta}{}^{d(a_{k-1})}
\end{eqnarray}
\begin{equation}
\delta_0 \Psi_{\mu\nu}{}^{(a_k)} = \partial_{[\mu} 
\xi_{\nu]}{}^{(a_k)} + \eta^{(a_k)}{}_{\mu\nu}, \qquad
\delta_0 \Omega_\mu{}^{(a_k),bc} = \partial_\mu
\eta^{(a_k),bc}
\end{equation}
where parameters $\xi$ and $\eta$ have the same properties on local
indices as $\Phi$ and $\Omega$, correspondingly.

Now, if we replace all derivatives in the Lagrangian and gauge
transformations by the covariant ones, we obtain non-invariance of the
form:
$$
(-1)^k \delta_0 {\cal L}_0 = \frac{2(k+2)(d+k-4)}{k} \kappa
\left\{ \phantom{|}^{\mu\nu}_{ab} \right\} [
\eta^{(a_k),ab} \Psi_{\mu\nu}{}^{(a_k)} -
\Omega_\mu{}^{(a_k),ab} \xi_\nu{}^{(a_k)} ]
$$
but in this case it is impossible to restore broken invariance
(without introduction of additional fields) mainly because there is no
covariant mass-like term for such field.

\subsection{Tensor $Y(k+1,k+1)$}

This case is also special and deserves separate consideration. Main
physical field now --- two form $R_{\mu\nu}{}^{(a_k),(b_k)}$ symmetric
on both groups of local indices, completely traceless on all local
indices and satisfying constraints $R_{\mu\nu}{}^{(a_k),(b_k)} =
R_{\mu\nu}{}^{(b_k),(a_k)}$ and $R_{\mu\nu}{}^{(a_k,b_1)(b_{k-1})} =
0$, while auxiliary field is a two form 
$\Omega_{\mu\nu}{}^{(a_k),(b_k),c}$. The Lagrangian and gauge
transformations for massless field have the form:
\begin{eqnarray}
{\cal L}_0 &=& - \frac{1}{2}
\left\{ \phantom{|}^{\mu\nu\alpha\beta}_{abcd} \right\} [
\Omega_{\mu\nu}{}^{a(a_{k-1}),b(b_{k-1}),e} 
\Omega_{\alpha\beta}{}^{c(a_{k-1}),d(b_{k-1}),e} + \nonumber \\
 && \qquad \qquad \quad + \frac{2}{k}
\Omega_{\mu\nu}{}^{(a_k),a(b_{k-1}),b}
\Omega_{\alpha\beta}{}^{(a_k),c(b_{k-1}),d} + \nonumber \\
 && + \left\{ \phantom{|}^{\mu\nu\alpha\beta\gamma}_{abcde} \right\}
\Omega_{\mu\nu}{}^{a(a_{k-1}),b(b_{k-1}),c} \partial_\alpha
R_{\beta\gamma}{}^{d(a_{k-1}),e(b_{k-1})}
\end{eqnarray}
\begin{equation}
\delta R_{\mu\nu}{}^{(a_k),(b_k)} = \partial_{[\mu}
\xi_{\nu]}{}^{(a_k),(b_k)} + \eta_{[\mu}{}^{(a_k),(b_k)}{}_{\nu]},
\qquad \delta \Omega_{\mu\nu}{}^{(a_k),(b_k),c} = \partial_{[\mu}
\eta_{\nu]}{}^{(a_k),(b_k),c}
\end{equation}

One of the peculiar features of this tensor is that it admits
deformation into $AdS$ space without introduction of any additional
fields. Indeed, non-invariance of the Lagrangian that appears if we
replace all derivatives by the covariant ones looks as follows:
$$
\delta_0 {\cal L}_0 = \frac{4(k+2)(d+k-5)}{k} \kappa
\left\{ \phantom{|}^{\mu\nu\alpha}_{abc} \right\} [
\Omega_{\mu\nu}{}^{(a_k),a(b_{k-1}),b}
\xi_\alpha{}^{(a_k),c(b_{k-1})} + 
\eta_\mu{}^{(a_k),a(b_{k-1}),b}
R_{\nu\alpha}{}^{(a_k),c(b_{k-1})} ]
$$
But the invariance can be restored by adding to the Lagrangian
mass-like term of the form:
\begin{equation}
{\cal L}_2 = a_{k,k}
\left\{ \phantom{|}^{\mu\nu\alpha\beta}_{abcd} \right\}
R_{\mu\nu}{}^{a(a_{k-1}),b(b_{l-1})}
R_{\alpha\beta}{}^{c(a_{k-1}),d(b_{l-1})}
\end{equation}
as well as appropriate corrections for gauge transformations:
\begin{eqnarray}
\delta_2 \Omega_{\mu\nu}{}^{(a_k),(b_k),c} &=& - 
\frac{2a_{k,k}}{(k+2)(d-4)}
 [ k e_{[\mu}{}^c \xi_{\nu]}{}^{(a_k),(b_k)} -
e_{[\mu}{}^{(a_1} \xi_{\nu]}{}^{a_{k-1})c,(b_k)} -
e_{[\mu}{}^{(b_1} \xi_{\nu]}{}^{(a_k),b_{k-1})c} - \nonumber \\
 && \qquad - \frac{1}{(d+k-5)} [
 ( g^{(a_1a_2} \xi_{[\mu,\nu]}{}^{a_{k-2})c,(b_k)} +
g^{(b_1b_2} \xi_{[\mu}{}^{(a_k),b_{k-2})c}{}_{\nu]} ) - \nonumber \\
 && \qquad \qquad \qquad \qquad
 - (k-1) ( g^{c(a_1} \xi_{[\mu,\nu]}{}^{a_{k-1}),(b_k)}
+ g^{c(b_1} \xi_{[\mu}{}^{(a_k),b_{k-1})}{}_{\nu]} ) + \nonumber \\
 && \qquad \qquad \qquad \qquad +
g^{(a_1(b_1} ( \xi_{[\mu,\nu]}{}^{a_{k-1}),b_{k-1})c} +
\xi_{[\mu}{}^{a_{k-1})c,b_{k-1})}{}_{\nu]} ) ] ]
\end{eqnarray}
provided $a_{k,k} = \frac{(k+2)}{2} (d+k-5) \kappa$.

\subsection{Tensor $Y(k+1,l+1)$}

Now we are ready to consider general case --- $Y(k+1,l+1)$, $k > l >
0$. Main physical field here --- two form 
$\Psi_{\mu\nu}{}^{(a_k),(b_l)}$ symmetric on both groups of local
indices, completely traceless on all local indices and satisfying a
constraint $\Psi_{\mu\nu}{}^{(a_k,b_1)(b_{l-1})} = 0$, while auxiliary
field is a two form $\Omega_{\mu\nu}{}^{(a_k),(b_l),c}$. The
Lagrangian and gauge transformations for massless field have the form:
\begin{eqnarray}
(-1)^{k+l} {\cal L}_0 &=& - \frac{1}{2}
\left\{ \phantom{|}^{\mu\nu\alpha\beta}_{abcd} \right\} [
\Omega_{\mu\nu}{}^{a(a_{k-1}),b(b_{l-1}),e} 
\Omega_{\alpha\beta}{}^{c(a_{k-1}),d(b_{l-1}),e} + \nonumber \\
 && \qquad \qquad \quad + \frac{1}{k}
\Omega_{\mu\nu}{}^{(a_k),a(b_{l-1}),b}
\Omega_{\alpha\beta}{}^{(a_k),c(b_{l-1}),d} + \nonumber \\
 && \qquad \qquad \quad + \frac{1}{l} 
\Omega_{\mu\nu}{}^{a(a_{k-1}),(b_l),b}
\Omega_{\alpha\beta}{}^{c(a_{k-1}),(b_l),d} ] + \nonumber \\
 && + \left\{ \phantom{|}^{\mu\nu\alpha\beta\gamma}_{abcde} \right\}
\Omega_{\mu\nu}{}^{a(a_{k-1}),b(b_{l-1}),c} \partial_\alpha
\Psi_{\beta\gamma}{}^{d(a_{k-1}),e(b_{l-1})}
\end{eqnarray}
\begin{equation}
\delta \Psi_{\mu\nu}{}^{(a_k),(b_l)} = \partial_{[\mu}
\xi_{\nu]}{}^{(a_k),(b_l)} + \eta_{[\mu}{}^{(a_k),(b_l)}{}_{\nu]},
\qquad \delta \Omega_{\mu\nu}{}^{(a_k),(b_l),c} = \partial_{[\mu}
\eta_{\nu]}{}^{(a_k),(b_l),c}
\end{equation}

Non-invariance of the Lagrangian that appears if we replace all
derivatives by the covariant ones looks as follows:
\begin{eqnarray*}
(-1)^{k+l} \delta_0 {\cal L}_0 &=& 2 \kappa
\left\{ \phantom{|}^{\mu\nu\alpha}_{abc} \right\} [
\frac{[(k+2)(d+k-4) + (d+l-6)]}{k}
\Omega_{\mu\nu}{}^{(a_k),a(b_{l-1}),b}
\xi_\alpha{}^{(a_k),c(b_{l-1})} + \\
 && \qquad   \qquad + \frac{(l+1)(d+l-6)}{l} 
\Omega_{\mu\nu}{}^{a(a_{k-1}),(b_l),b}
\xi_\alpha{}^{c(a_{k-1}),(b_l)} + \\
&& \qquad \qquad + \frac{[(k+2)(d+k-4) + (d+l-6)]}{k}
\eta_\mu{}^{(a_k),a(b_{l-1}),b}
\Psi_{\nu\alpha}{}^{(a_k),c(b_{l-1})} + \\
 && \qquad   \qquad + \frac{(l+1)(d+l-6)}{l}
\eta_\mu{}^{a(a_{k-1}),(b_l),b}
\Psi_{\nu\alpha}{}^{c(a_{k-1}),(b_l)} ]
\end{eqnarray*}
We could try to restore broken gauge invariance by adding mass-like
terms to the Lagrangian:
\begin{equation}
(-1)^{k+l} {\cal L}_2 = a_{k,l}
\left\{ \phantom{|}^{\mu\nu\alpha\beta}_{abcd} \right\}
\Psi_{\mu\nu}{}^{a(a_{k-1}),b(b_{l-1})}
\Psi_{\alpha\beta}{}^{c(a_{k-1}),d(b_{l-1})}
\end{equation}
as well as appropriate corrections for gauge transformations:
\begin{eqnarray}
\delta_2 \Omega_{\mu\nu}{}^{(a_k),(b_l),c} &=& - 
\frac{2 a_{k,l}}{(k+2)(l+1)(d-4)}
 [ (k+1) l e_{[\mu}{}^c \xi_{\nu]}{}^{(a_k),(b_l)} -
 l e_{[\mu}{}^{(a_1} \xi_{\nu]}{}^{a_{k-1})c,(b_l)} + \nonumber \\
&& \qquad \qquad \quad + 
e_{[\mu}{}^{(a_1} \xi_{\nu]}{}^{a_{k-1})(b_1,b_{l-1})c}
- (k+1) e_{[\mu}{}^{(b_1} \xi_{\nu]}{}^{(a_k),b_{l-1})c} + \dots ]
\label{ap1}
\end{eqnarray}
where dots stand for the terms which are necessary to make
transformations to be traceless (see Appendix), but this gives
additional variations of the form:
\begin{eqnarray*}
 - 4 (-1)^{k+l} a_{k,l} 
&\left\{ \phantom{|}^{\mu\nu\alpha}_{abc} \right\}& [
 \frac{1}{k} \eta_\mu{}^{(a_k),a(b_{l-1}),b}
\Psi_{\nu\alpha}{}^{(a_k),c(b_{l-1})} + \frac{1}{l}
\eta_\mu{}^{a(a_{k-1}),(b_l),b}
\Psi_{\nu\alpha}{}^{c(a_{k-1}),(b_l)} + \\
 && + \frac{1}{k} \Omega_{\mu\nu}{}^{(a_k),a(b_{l-1}),b}
\xi_\alpha{}^{(a_k),c(b_{l-1})} + \frac{1}{l}
\Omega_{\mu\nu}{}^{a(a_{k-1}),(b_l),b}
\xi_\alpha{}^{c(a_{k-1}),(b_l)} ]
\end{eqnarray*}
so that it is impossible to achieve a cancellation by adjusting the
only free parameter $a_{k,l}$.

\section{Massive tensor $Y(k+1,k+1)$}

We have already seen in the previous section that tensor $Y(k+1,k+1)$
turns out to be special and deserves separate consideration. In this
section we consider massive theory for such tensor. Recall that
frame-like gauge invariant formulation for massive tensors $Y(k+1,0)$
and $Y(k+1,1)$ has been constructed in \cite{Zin08b} and \cite{Zin08c}
correspondingly.

First of all, to construct gauge invariant formulation for massive
field we have to determine a set of additional Goldstone fields
required for such formulation. In general for each gauge symmetry of
main gauge field we have to introduce corresponding primary Goldstone
field. But these primary fields turn out to be gauge fields themselves
with their own gauge symmetries so we have to introduce secondary
fields and so on. In this, working with mixed symmetry (spin)-tensors
one has to take into account reducibility of their gauge
transformations. Let us illustrate the procedure on this particular
case. Our main field $Y(k+1,k+1)$ has one gauge symmetry with the
parameter corresponding to $Y(k+1,k)$ and this transformations are
reducible with the reducibility parameter $Y(k,k)$. So we introduce
primary field $Y(k+1,k)$. This field in turn has two gauge symmetries
with parameters $Y(k+1,k-1)$ and $Y(k,k)$ and reducibility parameter
$Y(k,k-1)$, thus taking into account reducibility of main field gauge
transformations it is enough to introduce one secondary field
$Y(k+1,k-1)$ only. It is not hard to see that the procedure stops at
the field $Y(k+1,0)$ corresponding to completely symmetric tensor and
having one gauge transformation with parameter $Y(k,0)$ only. Thus the
minimal set of fields necessary for gauge invariant description
contains the fields $Y(k+1,n)$ with $(0 \le n \le k+1)$.

In frame-like gauge invariant formalism for massive bosonic fields the
general structure of the Lagrangian looks as follows (schematically):
$$
{\cal L} \sim {\cal L}_0 + {\cal L}_1 + {\cal L}_2, \qquad
{\cal L}_0 \sim \Omega \Omega + \Omega \partial \Phi, \qquad
{\cal L}_1 \sim m \Omega \Phi, \qquad {\cal L}_2 \sim m^2 \Phi \Phi
$$
where ${\cal L}_0$ and ${\cal L}_2$ are kinetic and mass terms for all
fields involved, while ${\cal L}_1$ --- a set of cross terms mixing
different fields together (see below). Similarly, the general
structure of gauge transformations looks like:
$$
\delta \sim \delta_0 + \delta_1 + \delta_2, \qquad
\delta_0 \Phi \sim \partial \xi + \eta, \quad \delta_0 \Omega \sim
\partial \eta, \qquad \delta_1 \Phi \sim m \xi, \quad \delta_1 \Omega
\sim m \eta, \qquad \delta_2 \Omega \sim m^2 \xi
$$
Thus it is convenient to organize the variations of the Lagrangian by
the order of mass parameter $m$ (in a metric-like formulation this
corresponds to the number of derivatives in variations). Taking into
account that in flat Minkowski space our kinetic terms are already
gauge invariant $\delta_0 {\cal L}_0 = 0$, while in $(A)dS$ spaces
they give non-trivial contributions of order $m^2$ (due to
non-commutativity of covariant derivatives), the Lagrangian will be
gauge invariant provided:
$$
\delta_0 {\cal L}_1 + \delta_1 {\cal L}_0 = 0
$$
$$
\delta_0 {\cal L}_0 + \delta_1 {\cal L}_1 + \delta_0 {\cal L}_2 +
\delta_2 {\cal L}_0 = 0
$$
$$
\delta_1 {\cal L}_2 + \delta_2 {\cal L}_1 = 0
$$

Following this general structure we introduce kinetic and mass terms
for all fields involved:
\begin{equation}
{\cal L}_0 = {\cal L}_0(R_{\mu\nu}{}^{(a_k),(b_k)}) +
\sum_{n=0}^{k-1} {\cal L}_0(\Psi_{\mu\nu}{}^{(a_k),(b_n)}) +
{\cal L}_0(\Phi_\mu{}^{(a_k)})
\end{equation}
\begin{eqnarray}
{\cal L}_2 &=& {\cal L}_2(R_{\mu\nu}{}^{(a_k),(b_k)}) +
\sum_{n=1}^{k-1} {\cal L}_2(\Psi_{\mu\nu}{}^{(a_k),(b_{n})}) +
{\cal L}_2(\Phi_\mu{}^{(a_k)}) + \nonumber \\
 && + b_{k,1} \left\{ \phantom{|}^{\mu\nu\alpha}_{abc} \right\}
\Psi_{\mu\nu}{}^{a(a_{k-1}),b} \Phi_\alpha{}^{c(a_{k-1})}
\end{eqnarray}
where all massless Lagrangians and mass terms  (as well as initial
gauge transformations $\delta_0$ and $\delta_2$) are exactly the same
as in the previous section, but with ordinary derivatives replaced by
the covariant ones. Note that there is a possibility of mixing between
$Y(k+1,2)$ and $Y(k+1,0)$ in mass terms with the corresponding
corrections for gauge transformations of the form:
\begin{eqnarray}
\delta_2 \Omega_{\mu\nu}{}^{(a_k),bc} &=& 
\frac{b_{k,1}}{2(k+2)(d-3)(d-4)}
[ k e_{[\mu}{}^b e_{\nu]}{}^c \zeta^{(a_k)} -
e_{[\mu}{}^b e_{\nu]}{}^{(a_1} \zeta^{a_{k-1})c} +
e_{[\mu}{}^c e_{\nu]}{}^{(a_1} \zeta^{a_{k-1})b} + \nonumber \\
 && \qquad  + \frac{1}{(d+k-4)} [
2 g^{(a_1a_2} e_{[\mu}{}^{[b} \zeta_{\nu]}{}^{a_{k-2})c]} +
(k-1)  g^{(a_1[b} e_\mu{}^{c]} \zeta_{\nu]}{}^{a_{k-1})} - \nonumber
\\
 && \qquad \qquad \qquad \qquad -
 g^{(a_1[b} e_{[\mu}{}^{a_2} \zeta_{\nu]}{}^{a_{k-2})c]} ] ] \\
\delta_2 \omega_\mu{}^{a,(a_k)} &=& - b_{k,1} \xi_\mu{}^{(a_k),a}
\nonumber
\end{eqnarray}
while the mass term for the $Y(k+1,1)$ field is absent.

Our next and the most important task --- to construct a complete set
of cross terms ${\cal L}_1$ as well as corresponding corrections for
gauge transformations so that $\delta_0 {\cal L}_1 + \delta_1 
{\cal L}_0 = 0$. All our previous experience tells that it is enough
to introduce such cross terms for the nearest neighbours only, i.e.
main field with primary fields, primary with secondary and so on. Then
the structure of general massive theory looks as shown on Figure 1,
where each arrow corresponds to a cross term mixing two fields, while
parameters $d_{k,n}$ will be determined below.
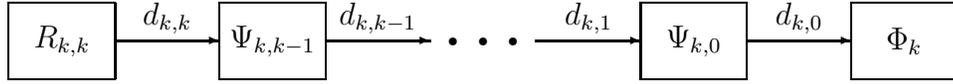
\begin{figure}[htb]
\begin{center}
\begin{picture}(146,12)
\put(10,1){\framebox(14,10)[]{$R_{k,k}$}}
\put(27,6){\makebox(8,6)[]{$d_{k,k}$}}
\put(38,1){\framebox(14,10)[]{$\Psi_{k,k-1}$}}
\put(55,6){\makebox(8,6)[]{$d_{k,k-1}$}}
\put(94,1){\framebox(14,10)[]{$\Psi_{k,0}$}}
\put(83,6){\makebox(8,6)[]{$d_{k,1}$}}
\put(122,1){\framebox(14,10)[]{$\Phi_k$}}
\put(111,6){\makebox(8,6)[]{$d_{k,0}$}}
\multiput(24,6)(28,0){4}{\vector(1,0){14}}
\multiput(69,6)(4,0){3}{\circle*{1}}
\end{picture}
\end{center}
\caption{General massive $Y(k+1,k+1)$ theory}
\end{figure}

Let us consider all possible cross terms in turn.

$Y(k+1,n+1) \Leftrightarrow Y(k+1,n)$. Most of the cross terms  belong
to this class except the $n=1$ and $n=0$ cases which have to be
considered separately. In this case additional terms for the
Lagrangian look as follows:
\begin{eqnarray}
(-1)^{k+n} {\cal L}_1 &= & d_{k,n}
\left\{ \phantom{|}^{\mu\nu\alpha\beta}_{abcd} \right\}
 [ \Omega_{\mu\nu}{}^{a(a_{k-1}),b(b_{n-1}),c}
\Psi_{\alpha\beta}{}^{d(a_{k-1}),(b_{n-1})} + \nonumber \\
 && \qquad \qquad \quad + 
\Omega_{\mu\nu}{}^{a(a_{k-1}),(b_{n-1}),b}
\Psi_{\alpha\beta}{}^{c(a_{k-1}),d(b_{n-1})} ]
\end{eqnarray}
In this, to compensate their non-invariance under the initial gauge
transformations we have to introduce corresponding corrections for
gauge transformations:
\begin{eqnarray}
\delta_1 \Psi_{\mu\nu}{}^{(a_k),(b_n)} &=& - 
\frac{d_{k,n}}{(k-n+2)(d+n-5)} [ 
(k-n+1) \xi_{[\mu}{}^{(a_k),(b_{n-1}} e_{\nu]}{}^{b_1)} - \nonumber \\
 && \qquad \qquad \qquad \qquad \qquad -
e_{[\nu}{}^{(a_1} \xi_{\mu]}{}^{a_{k-1})(b_1,b_{n-1})} + \dots ]
\nonumber \\
\delta_1 \Omega_{\mu\nu}{}^{(a_k),(b_n),c} &=&
\frac{d_{k,n}}{(k-n+2)(d+n-4)} [
(k-n+1) e_{[\mu}{}^{(b_1} \eta_{\nu]}{}^{(a_k),b_{n-1}),c} - \nonumber
\\
 && \qquad \qquad \qquad \qquad \qquad -
 e_{[\mu}{}^{(a_1} \eta_{\nu]}{}^{a_{k-1})(b_1,b_{n-1}),c} + \dots ]
\label{ap2} \\
\delta_1 \Psi_{\mu\nu}{}^{(a_k),(b_{n-1})} &=& - \frac{(n-1)}{n}
d_{k,n} \xi_{[\mu}{}^{(a_k),(b_{n-1})}{}_{\nu]} \nonumber \\
\delta_1 \Omega_{\mu\nu}{}^{(a_k),(b_{n-1}),c} &=& - d_{k,n} [
\eta_{[\mu}{}^{(a_k),(b_{n-1})}{}_{\nu]}{}^c + \frac{1}{n}
\eta_{[\mu}{}^{(a_k),(b_{n-1})c,}{}_{\nu]} ] \nonumber
\end{eqnarray}
where again dots stand for the terms which are necessary to make
variations traceless and complete form of these expressions can be
found in Appendix.

$Y(k+1,2) \Leftrightarrow Y(k+1,1)$ Here the additional terms for the
Lagrangian has the following form:
\begin{eqnarray}
(-1)^{k+1} {\cal L}_1 &=& d_{k,1} 
\left\{ \phantom{|}^{\mu\nu\alpha\beta}_{abcd} \right\}
\Omega_{\mu\nu}{}^{a(a_{k-1}),bc} \Psi_{\alpha\beta}{}^{d(a_{k-1})} +
\nonumber \\
 && + d_{k,1}
\left\{ \phantom{|}^{\mu\nu\alpha}_{abc} \right\} [ \frac{1}{k}
 \Omega_\mu{}^{(a_k),ab} \Psi_{\nu\alpha}{}^{(a_k),c} - 2
\Omega_\mu{}^{a(a_{k-1}),be} 
\Psi_{\nu\alpha}{}^{c(a_{k-1}),e} ]
\end{eqnarray}
while corresponding corrections for gauge transformations turn out to
be:
\begin{eqnarray}
\delta_1 \Psi_{\alpha\beta}{}^{(a_k),b} &=& - 
\frac{d_{k,1}}{(k+1)(d-4)}
[ k \xi_{[\alpha}{}^{(a_k)} e_{\beta]}{}^b 
- e_{[\beta}{}^{(a_1} \xi_{\alpha]}{}^{a_{k-1})b} + \nonumber \\
 && \qquad \qquad + \frac{1}{(d+k-3)} 
[ 2 g^{(a_1a_2} \xi_{[\alpha,\beta]}{}^{a_{k-2})b}
- (k-1) g^{b(a_1} \xi_{[\alpha,\beta]}{}^{a_{k-1})} ]  ] \nonumber \\
\delta_1 \Omega_{\mu\nu}{}^{(a_k),bc} &=& 
- \frac{k d_{k,1}}{2(k+1)^2 (d-4)} 
[ (k+1) \eta^{(a_k),[b}{}_{[\mu} e_{\nu]}{}^{c]} + 
e_{[\nu}{}^{(a_1} \eta^{a_{k-1})[b,c]}{}_{\mu]} + \nonumber \\
 && \quad + \frac{1}{(d+k-4)} [
2 g^{(a_1a_2} \eta_{[\mu}{}^{a_{k-2})[b,c]}{}_{\nu]}
- k g^{[b(a_1} \eta_{[\mu}{}^{a_{k-1}),c]}{}_{\nu]} 
- 2 g^{[b(a_1} \eta^{a_{k-1})c]}{}_{\mu\nu} ] ] + \nonumber \\
 && + \frac{d_{k,1}}{2(k+1)^2 (d+k-3)}
[ (k+1) e_{[\nu}{}^{(a_1}  \eta_{\mu]}{}^{a_{k-1}),bc}
- e_{[\nu}{}^{(a_1} \eta^{a_{k-1})[b,c]}{}_{\mu]} -  \\
 && \quad - \frac{1}{(d+k-4)} [
2 g^{(a_1a_2} \eta_{[\mu}{}^{a_{k-2})[b,c]}{}_{\nu]}
- k g^{[b(a_1} \eta_{[\mu}{}^{a_{k-1}),c]}{}_{\nu]} 
- 2 g^{[b(a_1} \eta^{a_{k-1})c]}{}_{\mu\nu} ] ] \nonumber \\
\delta_1 \Psi_{\alpha\beta}{}^{(a_k)} &=& - d_{k,1}
\xi_{[\alpha}{}^{(a_k)}{}_{\beta]}, \qquad
\delta_1 \Omega_\mu{}^{(a_k),bc} = - 2 d_{k,1}
\eta_\mu{}^{(a_k),bc} \nonumber
\end{eqnarray}
Note that in this (and only this) case there exist two independent
ways to obtain correct tensor structure for 
$\Omega_{\mu\nu}{}^{(a_k),bc}$  transformations\footnote{Author is
grateful to E. D. Skvortsov for pointing this out}.

$Y(k+1,1) \Leftrightarrow Y(k+1,0)$ These terms have already been
considered in \cite{Zin08c}, but we reproduce them here in our current
notations:
\begin{equation}
(-1)^k {\cal L}_1 = 
d_{k,0} [ \frac{1}{k} \left\{ \phantom{|}^{\mu\nu}_{ab} \right\}
\Omega_\mu{}^{(a_k),ab} \Phi_\nu{}^{(a_k)} + 
\left\{ \phantom{|}^{\mu\nu\alpha}_{abc} \right\}
\omega_\mu{}^{a,b(a_{k-1})} \Psi_{\nu\alpha}{}^{c(a_{k-1})} ]
\end{equation}
together with appropriate corrections for gauge transformations:
\begin{eqnarray}
\delta_1 \Psi_{\mu\nu}{}^{(a_k)} &=& - \frac{d_{k,0}}{2(k+2)(d+k-4)}
e_{[\mu}{}^{(a_1} \zeta_{\nu]}{}^{a_{k-1})} \nonumber \\
\delta_1 \Omega_\mu{}^{(a_k),bc} &=& - \frac{d_{k,0}}{2(k+2)(d-3)}
 [ (k+1) e_\mu{}^{[b} \chi^{c],(a_k)} + e_\mu{}^{(a_1}
\chi^{[b,c] a_{k-1})} + \\
 && - \frac{1}{(d+k-4)} [
2 g^{(a_1a_2} \chi^{[b,c]a_{k-2})}{}_\mu 
+ g^{(a_1[b} \chi_\mu{}^{c]a_{k-1})} 
+ k g^{(a_1[b} \chi^{c],a_{k-1})}{}_\mu ]  ] \nonumber \\
\delta_1 \Phi_\mu{}^{(a_k)} &=& - d_{k,0} \xi_\mu{}^{(a_k)},
\qquad \delta_1 \omega_\mu{}^{a,(a_k)} = - \frac{d_{k,0}}{2}
\eta^{(a_k),a}{}_\mu \nonumber
\end{eqnarray}

At this point we have a whole Lagrangian (i.e. a complete set of
kinetic, cross and mass terms) as well as complete set of gauge
transformations. In this, all parameters in gauge transformations are
expressed in terms of the Lagrangian ones $d_{k,n}$ and $a_{k,n}$ so
that all variations of order $m$ cancel $\delta_0 {\cal L}_1 +
\delta_1 {\cal L}_0 = 0$. Thus our next task --- to consider all
variations of order $m^2$ and $m^3$ and require their cancellation. We
will not give here these lengthy but straightforward calculations and
reproduce the final results only.

Recall that working with gauge invariant formulation for massive
fields it is natural to define massless limit as the one where all
additional Goldstone fields decouple from the main one. In the case at
hands this corresponds to the limit $d_{k,k}  \to 0$ and indeed in a
flat Minkowski space this parameter would be proportional to the mass.
As we have already mentioned in the Introduction, we will not try to
give any strict definition of what is mass in $(A)dS$ spaces. Instead,
we will simply use this parameter for the investigation of all special
limits that exist in such spaces. Let us denote $M^2 = (k-1)
d_{k,k}^2$. Then for the parameter $a_{k,k}$ determining mass term for
the main field we obtain:
$$
a_{k,k} = - \frac{(k+2)}{4k} [ M^2 - 2k(d+k-5) \kappa ]
$$
Moreover, all other mass terms turn out to be proportional to the main
one:
$$
a_{k,n} = \frac{k(d+k-4)}{n(d+n-4)} a_{k,k}, \qquad n \ge 1
$$
the only exceptions being:
$$
b_{k,1} = d_{k,1} d_{k,0}, \qquad
b_k = \frac{(d-2)}{(d-4)} d_{k,1}^2
$$

Also we obtain a number of recurrent relations on the parameters
$d_{k,n}$ which can be easily solved and give us:
$$
d_{k,n}^2 = \frac{(k-n+2)(d+k+n-4)}{2(n-1)(d+2n-4)}
[ M^2 - 2(k-n)(d+k+n-5) \kappa ], \qquad n \ge 2
$$
$$
d_{k,1}^2 = \frac{(k+1)(d+k-3)}{2(d-2)} 
[ M^2 - 2(k-1)(d+k-4) \kappa ]
$$
$$
d_{k,0}^2 = \frac{(k+2)(d+k-4)}{(d-4)}
[ M^2 - 2 k (d+k-5) \kappa ]
$$

Thus all the parameters in the Lagrangian and gauge transformations
are expressed in terms of one main parameter $M$ and we are ready to
analyze the results obtained. 

Let us begin with $AdS$ space. As we have already noted in the
previous section, massless tensor $Y(k+1,k+1)$ admits a deformation
into $AdS$ space without introduction of any additional fields. And
indeed, from the formulas given above we see that nothing prevent us
from considering massless limit $M \to 0$ where all additional fields
decouple from the main one as shown on Figure 2.
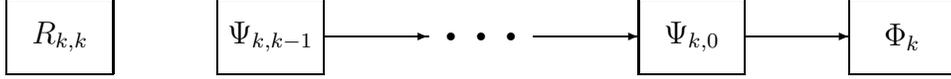
\begin{figure}[htb]
\begin{center}
\begin{picture}(146,12)
\put(10,1){\framebox(14,10)[]{$R_{k,k}$}}
\put(38,1){\framebox(14,10)[]{$\Psi_{k,k-1}$}}
\put(94,1){\framebox(14,10)[]{$\Psi_{k,0}$}}
\put(122,1){\framebox(14,10)[]{$\Phi_k$}}
\multiput(52,6)(28,0){3}{\vector(1,0){14}}
\multiput(69,6)(4,0){3}{\circle*{1}}
\end{picture}
\end{center}
\caption{Massless limit in $AdS$ space}
\end{figure}
In this, the remaining fields describe partially massless theory for
the $Y(k+1,k)$ tensor (see the discussion of such theories in the next
section).

Let us turn to the $dS$ space. First of all, from the expressions for
the parameters $d_{k,n}$ we see that there is a unitary forbidden
region $M^2 < 2k(d+k-5) \kappa$. At the boundary of this region we
obtain first partially massless theory where the last field $Y(k+1,0)$
decouples from the rest ones as shown on Figure 3.
\begin{figure}[htb]
\begin{center}
\begin{picture}(146,12)
\put(10,1){\framebox(14,10)[]{$R_{k,k}$}}
\put(38,1){\framebox(14,10)[]{$\Psi_{k,k-1}$}}
\put(94,1){\framebox(14,10)[]{$\Psi_{k,0}$}}
\put(122,1){\framebox(14,10)[]{$\Phi_k$}}
\multiput(24,6)(28,0){3}{\vector(1,0){14}}
\multiput(69,6)(4,0){3}{\circle*{1}}
\end{picture}
\end{center}
\caption{Unitary partially massless limit in $dS$ space}
\end{figure}
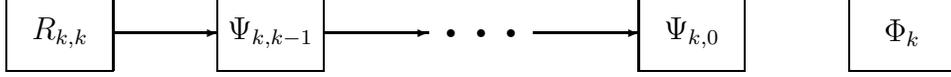
Note that at this point all mass parameters $a_{k,n}$ become equal to
zero. Inside the forbidden region we find a number of partially
massless limits which happens each time then one of the parameters
$d_{k,n}$ becomes equal to zero. In this case the whole system
decompose into two disconnected subsystems as shown on Figure 4.
\begin{figure}[htb]
\begin{center}
\begin{picture}(148,12)
\put(10,1){\framebox(14,10)[]{$R_{k,k}$}}
\put(24,6){\vector(1,0){10}}
\multiput(36,6)(3,0){3}{\circle*{1}}
\put(44,6){\vector(1,0){10}}
\put(54,1){\framebox(14,10)[]{$\Psi_{k,n}$}}

\put(80,1){\framebox(14,10)[]{$\Psi_{k,n-1}$}}
\put(94,6){\vector(1,0){10}}
\multiput(106,6)(3,0){3}{\circle*{1}}
\put(114,6){\vector(1,0){10}}
\put(124,1){\framebox(14,10)[]{$\Phi_k$}}
\end{picture}
\end{center}
\caption{Example of non-unitary partially massless limit in $dS$
space}
\end{figure}
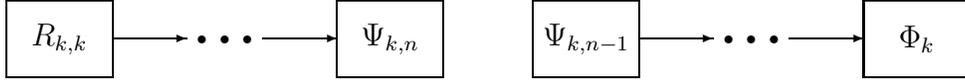
Both subsystems describe partially massless theories (for the
discussion of the second one see next section) however in this case we
have $d_{k,m}^2 > 0$ for $m > n$ while $d_{k,m}^2 < 0$ for $m < n$.

\section{Massive tensor $Y(k+1,l+1)$}

In this section we consider gauge invariant formulation for massive
$Y(k+1,l+1)$ tensor with $k \ne l$. This case will be more complicated
but our general strategy will be the same. First of all we have to
find a set of additional fields necessary for such description. Using
the fact that in general tensor $Y(m+1,n+1)$ has two gauge
transformations with parameters corresponding to $Y(m+1,n)$ and
$Y(m,n+1)$ and taking into account reducibility of these
transformations with parameter $Y(m,n)$, it is not hard to check that
we need $Y(m+1,n)$ with $l \le m \le k$, $0 \le n \le l+1$. All these
fields as well as parameters determining cross terms (see below) are
shown on Figure 5.
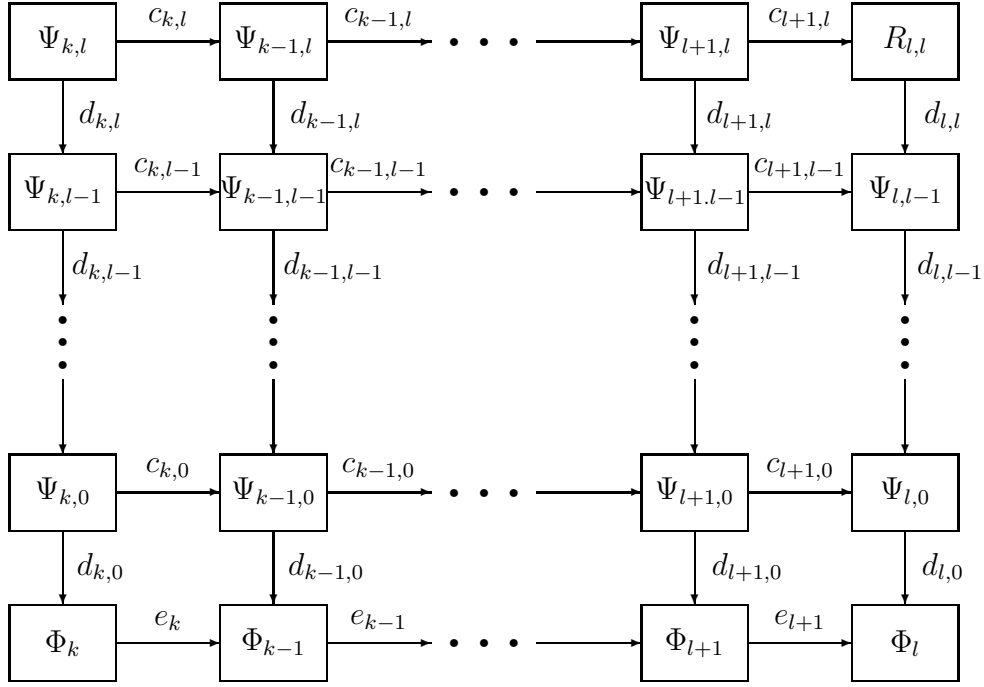
\begin{figure}[htb]
\begin{center}
\begin{picture}(146,92)
\put(10,1){\framebox(14,10)[]{$\Phi_k$}}
\put(27,6){\makebox(8,6)[]{$e_{k}$}}
\put(38,1){\framebox(14,10)[]{$\Phi_{k-1}$}}
\put(55,6){\makebox(8,6)[]{$e_{k-1}$}}
\put(94,1){\framebox(14,10)[]{$\Phi_{l+1}$}}
\put(111,6){\makebox(8,6)[]{$e_{l+1}$}}
\put(122,1){\framebox(14,10)[]{$\Phi_l$}}
\multiput(24,6)(28,0){4}{\vector(1,0){14}}
\multiput(69,6)(4,0){3}{\circle*{1}}
\put(10,21){\framebox(14,10)[]{$\Psi_{k,0}$}}
\put(27,26){\makebox(8,6)[]{$c_{k,0}$}}
\put(18,13){\makebox(8,6)[]{$d_{k,0}$}}
\put(38,21){\framebox(14,10)[]{$\Psi_{k-1,0}$}}
\put(55,26){\makebox(8,6)[]{$c_{k-1,0}$}}
\put(48,13){\makebox(8,6)[]{$d_{k-1,0}$}}
\put(94,21){\framebox(14,10)[]{$\Psi_{l+1,0}$}}
\put(111,26){\makebox(8,6)[]{$c_{l+1,0}$}}
\put(104,13){\makebox(8,6)[]{$d_{l+1,0}$}}
\put(122,21){\framebox(14,10)[]{$\Psi_{l,0}$}}
\put(130,13){\makebox(8,6)[]{$d_{l,0}$}}
\multiput(24,26)(28,0){4}{\vector(1,0){14}}
\multiput(69,26)(4,0){3}{\circle*{1}}
\put(10,61){\framebox(14,10)[]{$\Psi_{k,l-1}$}}
\put(27,66){\makebox(8,6)[]{$c_{k,l-1}$}}
\put(19,53){\makebox(8,6)[]{$d_{k,l-1}$}}
\put(38,61){\framebox(14,10)[]{$\Psi_{k-1,l-1}$}}
\put(55,66){\makebox(8,6)[]{$c_{k-1,l-1}$}}
\put(49,53){\makebox(8,6)[]{$d_{k-1,l-1}$}}
\put(94,61){\framebox(14,10)[]{$\Psi_{l+1.l-1}$}}
\put(111,66){\makebox(8,6)[]{$c_{l+1,l-1}$}}
\put(105,53){\makebox(8,6)[]{$d_{l+1,l-1}$}}
\put(122,61){\framebox(14,10)[]{$\Psi_{l,l-1}$}}
\put(131,53){\makebox(8,6)[]{$d_{l,l-1}$}}
\multiput(24,66)(28,0){4}{\vector(1,0){14}}
\multiput(69,66)(4,0){3}{\circle*{1}}
\put(10,81){\framebox(14,10)[]{$\Psi_{k,l}$}}
\put(27,86){\makebox(8,6)[]{$c_{k,l}$}}
\put(18,73){\makebox(8,6)[]{$d_{k,l}$}}
\put(38,81){\framebox(14,10)[]{$\Psi_{k-1,l}$}}
\put(55,86){\makebox(8,6)[]{$c_{k-1,l}$}}
\put(48,73){\makebox(8,6)[]{$d_{k-1,l}$}}
\put(94,81){\framebox(14,10)[]{$\Psi_{l+1,l}$}}
\put(111,86){\makebox(8,6)[]{$c_{l+1,l}$}}
\put(103,73){\makebox(8,6)[]{$d_{l+1,l}$}}
\put(122,81){\framebox(14,10)[]{$R_{l,l}$}}
\put(130,73){\makebox(8,6)[]{$d_{l,l}$}}
\multiput(24,86)(28,0){4}{\vector(1,0){14}}
\multiput(69,86)(4,0){3}{\circle*{1}}
\multiput(17,21)(0,20){4}{\vector(0,-1){10}}
\multiput(17,43)(0,3){3}{\circle*{1}}
\multiput(45,21)(0,20){4}{\vector(0,-1){10}}
\multiput(45,43)(0,3){3}{\circle*{1}}
\multiput(101,21)(0,20){4}{\vector(0,-1){10}}
\multiput(101,43)(0,3){3}{\circle*{1}}
\multiput(129,21)(0,20){4}{\vector(0,-1){10}}
\multiput(129,43)(0,3){3}{\circle*{1}}
\end{picture}
\end{center}
\caption{General massive $Y(k+1,l+1)$ theory}
\end{figure}

First of all we introduce kinetic and mass terms for all fields
involved:
\begin{equation}
{\cal L}_0 = \sum_{m=l}^k \sum_{n=0}^l {\cal L}_0
(\Psi_{\mu\nu}{}^{(a_m),(b_n)}) + \sum_{m=l}^k 
{\cal L}_0 (\Phi_\mu{}^{(a_m)})
\end{equation}
\begin{equation}
{\cal L}_2 = \sum_{m=l}^k \sum_{n=0}^l {\cal L}_2 
(\Psi_{\mu\nu}{}^{(a_m),(b_n)}) + \sum_{m=l}^k [
b_{m,1} \left\{ \phantom{|}^{\mu\nu\alpha}_{abc} \right\}
\Psi_{\mu\nu}{}^{a(a_{m-1}),b} \Phi_\alpha{}^{c(a_{m-1})} +
{\cal L}_2 (\Phi_\mu{}^{(a_m)}) ]
\end{equation}
where all kinetic and mass terms (as well as initial gauge
transformations $\delta_0$ and $\delta_2$) are the same as in Section
1 but with the ordinary derivatives replaced by the covariant ones and
we again take into account the possibility of mixing for $Y(m+1,2)$
and $Y(m+1,0)$ tensors.

Our next and important task --- to determine a set of cross terms
mixing these fields together as well as corresponding corrections for
gauge transformations. As in all previously considered cases it is
enough to introduce such terms for the nearest neighbours only. For
the case at hands it means that general $Y(m+1,n+1)$ tensor will have
cross terms with four nearest fields only as shown on Figure 6.
\begin{figure}[htb]
\begin{center}
\begin{picture}(90,52)
\put(38,1){\framebox(14,10)[]{$\Psi_{m,n-1}$}}
\put(10,21){\framebox(14,10)[]{$\Psi_{m+1,n}$}}
\put(24,26){\vector(1,0){14}}
\put(27,26){\makebox(8,6)[]{$c_{m+1,n}$}}
\put(38,21){\framebox(14,10)[]{$\Psi_{m,n}$}}
\put(52,26){\vector(1,0){14}}
\put(55,26){\makebox(8,6)[]{$c_{m,n}$}}
\put(45,21){\vector(0,-1){10}}
\put(47,13){\makebox(8,6)[]{$d_{m,n}$}}
\put(66,21){\framebox(14,10)[]{$\Psi_{m-1,n}$}}
\put(38,41){\framebox(14,10)[]{$\Psi_{m,n+1}$}}
\put(45,41){\vector(0,-1){10}}
\put(48,33){\makebox(8,6)[]{$d_{m,n+1}$}}
\end{picture}
\end{center}
\caption{Illustration on possible cross terms}
\end{figure}
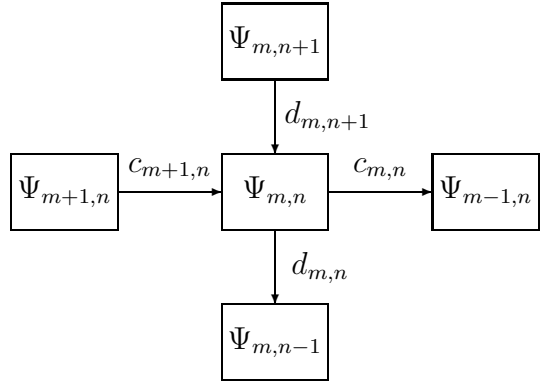
The cross terms corresponding to vertical arrows with parameters
$d_{m,n}$ have already been considered in the previous section so we
will not repeat them here. Let us consider cross terms corresponding
to horizontal arrows with parameters $c_{m,n}$.

$Y(m+1,n+1) \Leftrightarrow Y(m,n+1)$ In this case additional terms
for the Lagrangian can be written as follows:
\begin{eqnarray}
(-1)^{m+n} {\cal L}_1 &=& c_{m,n}
\left\{ \phantom{|}^{\mu\nu\alpha\beta}_{abcd} \right\} [
\Omega_{\mu\nu}{}^{a(a_{m-1}),b(b_{n-1}),c}
\Psi_{\alpha\beta}{}^{(a_{m-1}),d(b_{n-1})} + \nonumber \\
 && \qquad \qquad \quad +
\Omega_{\mu\nu}{}^{(a_{m-1}),a(b_{n-1}),b}
\Psi_{\alpha\beta}{}^{c(a_{m-1}),d(b_{n-1})} ]
\end{eqnarray}
while corresponding corrections for gauge transformations which are
necessary to compensate their non-invariance under the initial gauge
transformations look like (for complete expressions see Appendix):
\begin{eqnarray}
\delta_1 \Psi_{\mu\nu}{}^{(a_m),(b_n)} &=&
 - \frac{c_{m,n}}{(d+m-4)} [
e_{[\mu}{}^{(a_1} \xi_{\nu]}{}^{a_{m-1}),(b_n)} + \dots ]
\nonumber \\
\delta_1 \Omega_{\mu\nu}{}^{(a_m),(b_n),c} &=& - 
\frac{c_{m,n}}{(d+m-3)} [
 e_{[\mu}{}^{(a_1} \eta_{\nu]}{}^{a_{m-1}),(b_n),c} + \dots ]
\nonumber \\
\delta_1 \Psi_{\mu\nu}{}^{(a_{m-1}),(b_n)} &=&
\frac{(m-1)c_{m,n}}{(m+1)}  [ \xi_{[\mu,\nu]}{}^{(a_{m-1}),(b_n)}
+ \frac{1}{(m-n+1)}
\xi_{[\mu}{}^{(a_{m-1})(b_1,b_{n-1})}{}_{\nu]} ]  \label{ap3} \\
\delta_1 \Omega_{\mu\nu}{}^{(a_{m-1}),(b_n),c} &=& 
\frac{(m-1)c_{m,n}}{m}  [
\eta_{[\mu,\nu]}{}^{(a_{m-1}),(b_n),c} + \frac{1}{(m-n+1)}
\eta_{[\mu}{}^{(a_{m-1})(b_1,b_{n-1})}{}_{\nu]}{}^{c} + \nonumber \\
 && \qquad + \frac{1}{(m+1)} \eta_{[\mu}{}^{(a_{m-1})c,(b_n)}{}_{\nu]}
+ \frac{1}{(m+1)(m-n+1)}
\eta_{[\mu}{}^{(a_{m-1})(b_1,b_{n-1})c}{}_{\nu]} ] \nonumber
\end{eqnarray}

Once again there are two cases that have to be considered separately.

$Y(m+1,1) \Leftrightarrow Y(m,1)$ Here cross terms have the form:
\begin{equation}
(-1)^m {\cal L}_1 = c_{m,0}
\left\{ \phantom{|}^{\mu\nu\alpha}_{abc} \right\} [
\Omega_\mu{}^{a(a_{m-1}),bc} \Psi_{\nu\alpha}{}^{(a_{m-1})} +
\Omega_\mu{}^{(a_{m-1}),ab}  \Psi_{\nu\alpha}{}^{c(a_{m-1})} ]
\end{equation}
while corresponding corrections for gauge transformations turn out to
be:
\begin{eqnarray}
\delta_1 \Psi_{\mu\nu}{}^{(a_m)} &=& - \frac{c_{m,0}}{(d+m-4)}
[ e_{[\mu}{}^{(a_1} \xi_{\nu]}{}^{a_{m-1})} + \frac{2}{(d+2m-4)}
g^{(a_1a_2} \xi_{[\mu,\nu]}{}^{a_{m-2})}], \nonumber \\
\delta_1 \Omega_\mu{}^{(a_m),ab} &=& - \frac{c_{m,0}}{(d+m-2)}
[ e_\mu{}^{(a_1} \eta^{a_{m-1}),ab} - \frac{2}{(d+2m-4)}
g^{(a_1a_2} \eta_\mu{}^{a_{m-2}),ab} - \nonumber \\
 && - \frac{2}{(d+m-4)(d+2m-4)}
g^{(a_1a_2} \eta^{a_{m-2})[a,b]}{}_\mu +
\frac{1}{(d+m-4)} g^{(a_1[a} \eta^{a_{m-1}),b]}{}_\mu
 ] \nonumber  \\
\delta_1 \Psi_{\mu\nu}{}^{(a_{m-1})} &=&\frac{(m-1) c_{m,0}}{(m+1)} 
\xi_{[\mu,\nu]}{}^{(a_{m-1})} \\
\delta_1 \Omega_\mu{}^{(a_{m-1}),ab} &=& - c_{m,0}
[ \eta_\mu{}^{(a_{m-1}),ab} - \frac{1}{(m+1)}
\eta^{(a_{m-1})[a,b]}{}_\mu ] \nonumber
\end{eqnarray}

$Y(m+1,0) \Leftrightarrow Y(m,0)$ At last we need the following cross
terms:
\begin{equation}
(-1)^m {\cal L}_1 = e_m \left\{ \phantom{|}^{\mu\nu}_{ab} \right\} [
\omega_\mu{}^{a,b(a_{m-1})} \Phi_\nu{}^{(a_{m-1})} +
\omega_\mu{}^{a,(a_{m-1})} \Phi_\nu{}^{b(a_{m-1})} ]
\end{equation}
and corresponding corrections for gauge transformations:
\begin{eqnarray}
\delta_1 \Phi_\mu{}^{(a_m)} &=& - \frac{e_m}{2(d+m-3)}
[ e_\mu{}^{(a_1} \zeta^{a_{m-1})}  - \frac{2}{(d+2m-4)} g^{(a_1a_2}
\zeta_\mu{}^{a_{m-2})} ] \nonumber \\
\delta_1 \omega_\mu{}^{a,(a_m)} &=& - \frac{e_m}{2(d+m-2)}
[ \chi^{a,(a_{m-1}} e_\mu{}^{a_1)}  - \frac{2}{(d+2m-4)}
\chi^{a,(a_{k-2}}{}_\mu g^{a_1a_2)} + \nonumber \\
 && + \frac{2}{(d+m-3)(d+2m-4)}
g^{(a_1a_2} \chi_\mu{}^{a_{m-2})a} - 
\frac{1}{(d+m-3)} g^{a(a_1} \chi_\mu{}^{a_{m-1})}
] \nonumber \\
\delta_1 \Phi_\mu{}^{(a_{m-1})} &=& - \frac{(m-1) e_m}{2m}
\zeta_\mu{}^{(a_{m-1})} \\
\delta_1 \omega_\mu{}^{a,(a_{m-1})} &=& - \frac{e_m}{2}
[ \chi^{a,(a_{m-1})}{}_\mu + \frac{1}{m} \chi_\mu{}^{a(a_{m-1})} ]
\nonumber
\end{eqnarray}

At this point we have a whole Lagrangian (i.e. a complete set of
kinetic, cross and mass terms) as well as complete set of gauge
transformations. In this, all parameters in gauge transformations are
expressed in terms of the Lagrangian ones $d_{m,n}$, $c_{m,n}$ and
$a_{m,n}$ so that all variations of order $m$ cancel $\delta_0 {\cal
L}_1 + \delta_1 {\cal L}_0 = 0$. Thus our next task --- to consider
all variations of order $m^2$ and $m^3$ and require their
cancellation. Again we will not give here these very lengthy but
straightforward calculations and reproduce our final results only.

Our main tensor $Y(k+1,l+1)$ has two gauge transformations and as a
result there are two parameters $d_{k,l}$ and $c_{k,l}$ determining
its mixing with two primary fields $Y(k+1,l)$ and $Y(k,l+1)$.
According to our usual definition massless limit requires that both
$d_{k,l} \to 0$ and $c_{k,l} \to 0$ simultaneously. But this is
possible in a flat Minkowski space only (where both parameters are
proportional to mass). Again we will not insist on any definition of
what is mass in $(A)dS$ spaces and simply use one of these parameters
as our main one. Let us introduce a notation 
$M^2 = \frac{(k-l+1)(l-1)}{(k-l+2)} d_{k,l}^2$.  Then we obtain a
following expression for the parameter $a_{k,l}$ determining a mass
term for the main field:
$$
a_{k,l} = - \frac{(k+2)(l+1)}{2l(k+1)} [ M^2 - l (d+l-5) \kappa ]
$$
Moreover, all other mass terms turn out to be proportional to this
main one:
$$
a_{m,n} = \frac{l(k+1)(d+k-3)(d+l-4)}{n(m+1)(d+m-3)(d+n-4)} a_{k,l}
$$
the only exceptions being:
$$
b_{m,1} = d_{m,1} d_{m,0}, \qquad
b_m = \frac{(d-2)}{(d-4)} d_{m,1}^2
$$
Further we obtain the following expression for parameters $c_{m,l}$
corresponding to the topmost row on Figure 5:
$$
c_{m,l}^2 = \frac{(k-m+1)(d+k+m-2)}{(m-1)(d+2m-2)}
[ M^2 + (m-l+1) (d+m+l-4) \kappa ]
$$
as well as parameters $d_{k,n}$ corresponding to the leftmost column:
$$
d_{k,n}^2 = \frac{(l-n+1)(k-n+2)(d+l+n-4)}{(k-n+1)(n-1)(d+2n-4)}
[ M^2 - (l-n) (d+n+l-5) \kappa ], \qquad n \ge 2
$$
$$
d_{k,1}^2 = \frac{l(k+1)(d+l-3)}{k(d-2)} 
[ M^2 - (l-1) (d+l-4) \kappa ]
$$
$$
d_{k,0}^2 = \frac{2(k+2)(l+1)(d+l-4)}{(k+1)(d-4)}
[ M^2 - l (d+l-5) \kappa ]
$$
It is very important (and this gives a nice check for all lengthy
calculations) that all parameters $d_{m,n}$ corresponding to the same
row on Figure 5 turn out to be proportional to the leftmost one
$d_{k,n}$:
$$
d_{m,n}^2 = \frac{(k-n+1)(d+k+n-3)}{(m-n+1)(d+m+n-3)} d_{k,n}^2
$$
Similarly, all parameters $c_{m,n}$ and $e_m$ corresponding to the
same column turn out to be proportional to the topmost one $c_{m,l}$:
$$
c_{m,n}^2 = \frac{(m-l)(d+m+l-3)}{(m-n)(d+m+n-3)} c_{m,l}^2
$$
$$
e_m^2 = \frac{4(m-l)(d+m+l-3)}{(m+1)(d+m-4)} c_{m,l}^2
$$

Thus all the parameters in the Lagrangian and gauge transformations
are expressed in terms of one main parameter $M$ and we are ready to
investigate all special limits. Let us start with $AdS$ space ($\kappa
< 0$). From the expression for the parameters $c_{m,l}$ above we see
that in this case we have unitary forbidden region $M^2 < -
(k-l+1)(d+k+l-4) \kappa$. At the boundary of this region we find a
most interesting (and the only unitary) limit when $c_{k,l}$ (and
hence all $c_{m,l}$ and $e_l$) becomes equal to zero. In this case the
whole system decomposes into two disconnected subsystems as shown on 
Figure 7.
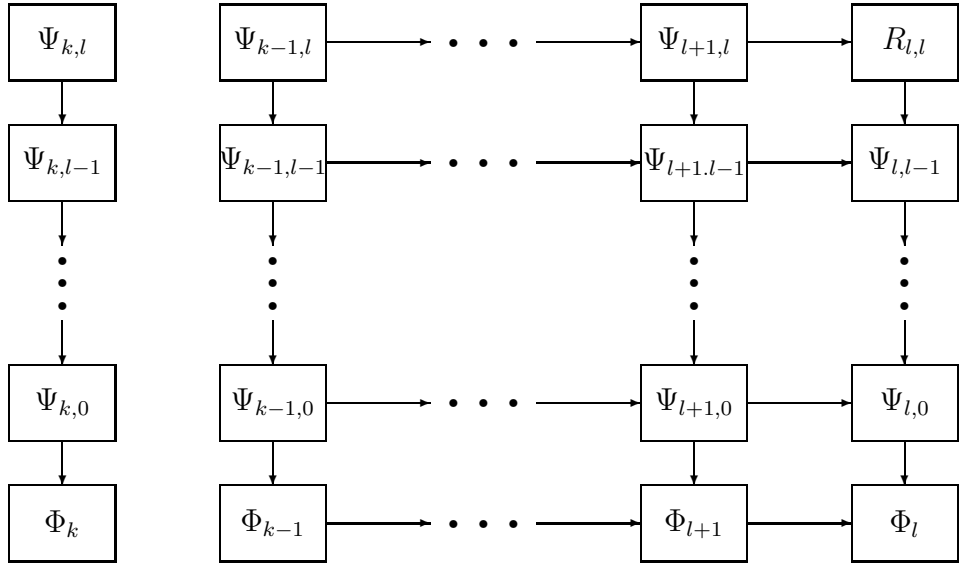
\begin{figure}[htb]
\begin{center}
\begin{picture}(130,77)
\put(1,1){\framebox(14,10)[]{$\Phi_k$}}
\put(29,1){\framebox(14,10)[]{$\Phi_{k-1}$}}
\put(85,1){\framebox(14,10)[]{$\Phi_{l+1}$}}
\put(113,1){\framebox(14,10)[]{$\Phi_l$}}
\multiput(43,6)(28,0){3}{\vector(1,0){14}}
\multiput(60,6)(4,0){3}{\circle*{1}}

\put(1,17){\framebox(14,10)[]{$\Psi_{k,0}$}}
\put(29,17){\framebox(14,10)[]{$\Psi_{k-1,0}$}}
\put(85,17){\framebox(14,10)[]{$\Psi_{l+1,0}$}}
\put(113,17){\framebox(14,10)[]{$\Psi_{l,0}$}}
\multiput(43,22)(28,0){3}{\vector(1,0){14}}
\multiput(60,22)(4,0){3}{\circle*{1}}

\put(1,49){\framebox(14,10)[]{$\Psi_{k,l-1}$}}
\put(29,49){\framebox(14,10)[]{$\Psi_{k-1,l-1}$}}
\put(85,49){\framebox(14,10)[]{$\Psi_{l+1.l-1}$}}
\put(113,49){\framebox(14,10)[]{$\Psi_{l,l-1}$}}
\multiput(43,54)(28,0){3}{\vector(1,0){14}}
\multiput(60,54)(4,0){3}{\circle*{1}}

\put(1,65){\framebox(14,10)[]{$\Psi_{k,l}$}}
\put(29,65){\framebox(14,10)[]{$\Psi_{k-1,l}$}}
\put(85,65){\framebox(14,10)[]{$\Psi_{l+1,l}$}}
\put(113,65){\framebox(14,10)[]{$R_{l,l}$}}
\multiput(43,70)(28,0){3}{\vector(1,0){14}}
\multiput(60,70)(4,0){3}{\circle*{1}}

\multiput(8,17)(0,16){4}{\vector(0,-1){6}}
\multiput(8,35)(0,3){3}{\circle*{1}}
\multiput(36,17)(0,16){4}{\vector(0,-1){6}}
\multiput(36,35)(0,3){3}{\circle*{1}}
\multiput(92,17)(0,16){4}{\vector(0,-1){6}}
\multiput(92,35)(0,3){3}{\circle*{1}}
\multiput(120,17)(0,16){4}{\vector(0,-1){6}}
\multiput(120,35)(0,3){3}{\circle*{1}}
\end{picture}
\end{center}
\caption{Unitary (partially) massless limit in $AdS$ space}
\end{figure}
A set of fields in the leftmost column describes partially massless
theory for the $Y(k+1,l+1)$ tensor. Note that from the anti-de Sitter
group point of view it corresponds to irreducible representation with
minimum number of degrees of freedom so such theory can be called
massless though in a flat limit it decomposes into a number of
massless representations of Lorentz group. Any way, this shows a
minimum number of additional fields which are necessary to deform
massless $Y(k+1,l+1)$ theory into $AdS$ space.  The pattern
corresponds to cutting the boxes from the second row of Young tableau
until we end up with diagram having one row only in complete
agreement with general discussion in \cite{BMV00}. At the same time
the rest fields describe massive theory for $Y(k,l+1)$ tensor.

Inside the unitary forbidden region we find a number of non-unitary
partially massless limits. They arise each time when one of the
$c_{m,l}$ (and hence all the parameters $c_{m,n}$ and $e_m$) becomes
equal to zero. In this case the whole system also decomposes into two
disconnected subsystems as shown on Figure 8.
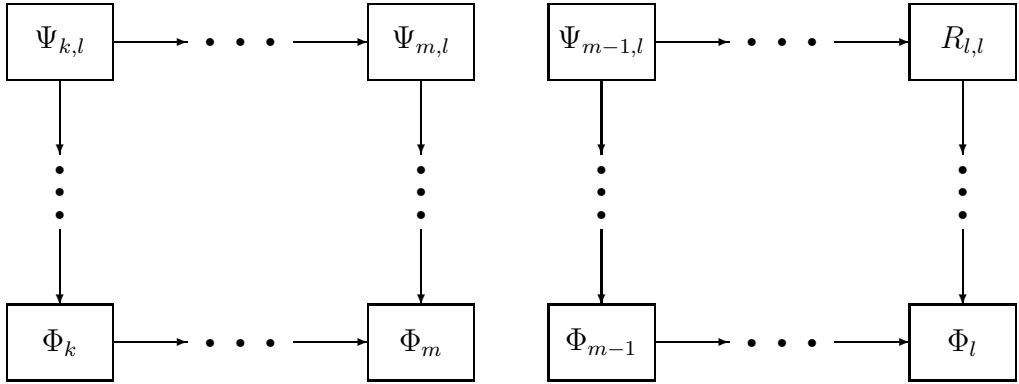
\begin{figure}[htb]
\begin{center}
\begin{picture}(137,52)
\put(1,1){\framebox(14,10)[]{$\Phi_k$}}
\put(8,21){\vector(0,-1){10}}
\multiput(8,23)(0,3){3}{\circle*{1}}
\put(8,41){\vector(0,-1){10}}
\put(1,41){\framebox(14,10)[]{$\Psi_{k,l}$}}
\put(15,6){\vector(1,0){10}}
\multiput(28,6)(4,0){3}{\circle*{1}}
\put(39,6){\vector(1,0){10}}
\put(49,1){\framebox(14,10)[]{$\Phi_m$}}
\put(15,46){\vector(1,0){10}}
\multiput(28,46)(4,0){3}{\circle*{1}}
\put(39,46){\vector(1,0){10}}
\put(49,41){\framebox(14,10)[]{$\Psi_{m,l}$}}
\put(56,21){\vector(0,-1){10}}
\multiput(56,23)(0,3){3}{\circle*{1}}
\put(56,41){\vector(0,-1){10}}

\put(73,1){\framebox(14,10)[]{$\Phi_{m-1}$}}
\put(80,21){\vector(0,-1){10}}
\multiput(80,23)(0,3){3}{\circle*{1}}
\put(80,41){\vector(0,-1){10}}
\put(73,41){\framebox(14,10)[]{$\Psi_{m-1,l}$}}
\put(87,6){\vector(1,0){10}}
\multiput(100,6)(4,0){3}{\circle*{1}}
\put(111,6){\vector(1,0){10}}
\put(121,1){\framebox(14,10)[]{$\Phi_l$}}
\put(87,46){\vector(1,0){10}}
\multiput(100,46)(4,0){3}{\circle*{1}}
\put(111,46){\vector(1,0){10}}
\put(121,41){\framebox(14,10)[]{$R_{l,l}$}}
\put(128,21){\vector(0,-1){10}}
\multiput(128,23)(0,3){3}{\circle*{1}}
\put(128,41){\vector(0,-1){10}}
\end{picture}
\end{center}
\caption{Example of non-unitary partially massless limit in $AdS$
space}
\end{figure}
Fields in the left block on Figure 8 describes partially massless
theory (this time really partially massless even from the anti-de
Sitter group point of view), while fields in the right block give
massive theory for $Y(m,l+1)$ tensor. Note however that in this case
we have $c_{p,l}^2 < 0$ for all $m < p \le k$.

Let us turn to the $dS$ space ($\kappa > 0$). From the expression for
the parameters $d_{k,n}$ above we see that in this case we also have
unitary forbidden region $M^2 < l(d+l-5) \kappa$. At the boundary of
this region we have $d_{k,0} = 0$ (and hence all $d_{m,0} = 0$) and we
find a first partially massless limit. Here the whole system also
decomposes into two disconnected subsystems (but this time diagram
splits vertically) as shown on Figure 9.
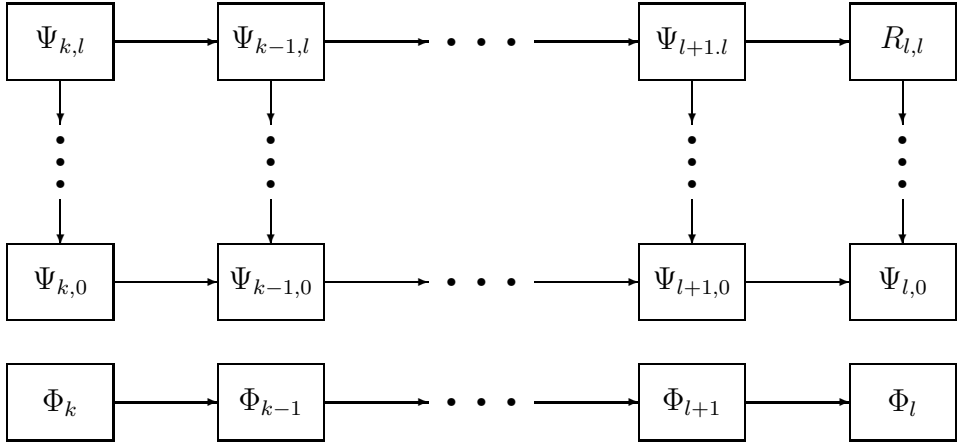
\begin{figure}[htb]
\begin{center}
\begin{picture}(130,62)
\put(1,1){\framebox(14,10)[]{$\Phi_k$}}
\put(29,1){\framebox(14,10)[]{$\Phi_{k-1}$}}
\put(85,1){\framebox(14,10)[]{$\Phi_{l+1}$}}
\put(113,1){\framebox(14,10)[]{$\Phi_l$}}
\multiput(15,6)(28,0){4}{\vector(1,0){14}}
\multiput(60,6)(4,0){3}{\circle*{1}}

\put(1,17){\framebox(14,10)[]{$\Psi_{k,0}$}}
\put(29,17){\framebox(14,10)[]{$\Psi_{k-1,0}$}}
\put(85,17){\framebox(14,10)[]{$\Psi_{l+1,0}$}}
\put(113,17){\framebox(14,10)[]{$\Psi_{l,0}$}}
\multiput(15,22)(28,0){4}{\vector(1,0){14}}
\multiput(60,22)(4,0){3}{\circle*{1}}

\put(1,49){\framebox(14,10)[]{$\Psi_{k,l}$}}
\put(29,49){\framebox(14,10)[]{$\Psi_{k-1,l}$}}
\put(85,49){\framebox(14,10)[]{$\Psi_{l+1.l}$}}
\put(113,49){\framebox(14,10)[]{$R_{l,l}$}}
\multiput(15,54)(28,0){4}{\vector(1,0){14}}
\multiput(60,54)(4,0){3}{\circle*{1}}

\multiput(8,33)(0,16){2}{\vector(0,-1){6}}
\multiput(8,35)(0,3){3}{\circle*{1}}
\multiput(36,33)(0,16){2}{\vector(0,-1){6}}
\multiput(36,35)(0,3){3}{\circle*{1}}
\multiput(92,33)(0,16){2}{\vector(0,-1){6}}
\multiput(92,35)(0,3){3}{\circle*{1}}
\multiput(120,33)(0,16){2}{\vector(0,-1){6}}
\multiput(120,35)(0,3){3}{\circle*{1}}
\end{picture}
\end{center}
\caption{Unitary partially massless limit in $dS$ space}
\end{figure}
Contrary to what we have seen in $AdS$ space both subsystems describe
partially massless theories for $Y(k+1,l+1)$ and $Y(k+1,0)$ tensors
correspondingly.

Inside the forbidden region we find a number of
partially massless limits. The most interesting one arises when
$d_{k,l} = 0$ (and hence all $d_{m,l} = 0$). In this case fields in
the upper row decouple from the rest ones as shown on Figure 10.
\begin{figure}[htb]
\begin{center}
\begin{picture}(118,72)
\put(10,1){\framebox(14,10)[]{$\Phi_k$}}
\put(17,21){\vector(0,-1){10}}
\multiput(17,23)(0,3){3}{\circle*{1}}
\put(17,41){\vector(0,-1){10}}
\put(10,41){\framebox(14,10)[]{$\Psi_{k,l-1}$}}
\put(24,6){\vector(1,0){14}}
\multiput(41,6)(4,0){3}{\circle*{1}}
\put(52,6){\vector(1,0){14}}
\put(66,1){\framebox(14,10)[]{$\Phi_l$}}
\put(24,46){\vector(1,0){14}}
\multiput(41,46)(4,0){3}{\circle*{1}}
\put(52,46){\vector(1,0){14}}
\put(66,41){\framebox(14,10)[]{$\Psi_{l,l-1}$}}
\put(73,21){\vector(0,-1){10}}
\multiput(73,23)(0,3){3}{\circle*{1}}
\put(73,41){\vector(0,-1){10}}

\put(10,61){\framebox(14,10)[]{$\Psi_{k,l}$}}
\put(24,66){\vector(1,0){14}}
\multiput(41,66)(4,0){3}{\circle*{1}}
\put(52,66){\vector(1,0){14}}
\put(66,61){\framebox(14,10)[]{$R_{l,l}$}}
\put(94,1){\framebox(14,10)[]{$\Phi_{l-1}$}}
\put(94,41){\framebox(14,10)[]{$R_{l-1,l-1}$}}
\put(101,21){\vector(0,-1){10}}
\multiput(101,23)(0,3){3}{\circle*{1}}
\put(101,41){\vector(0,-1){10}}
\end{picture}
\end{center}
\caption{(Partially) massless limit in $dS$ space}
\end{figure}
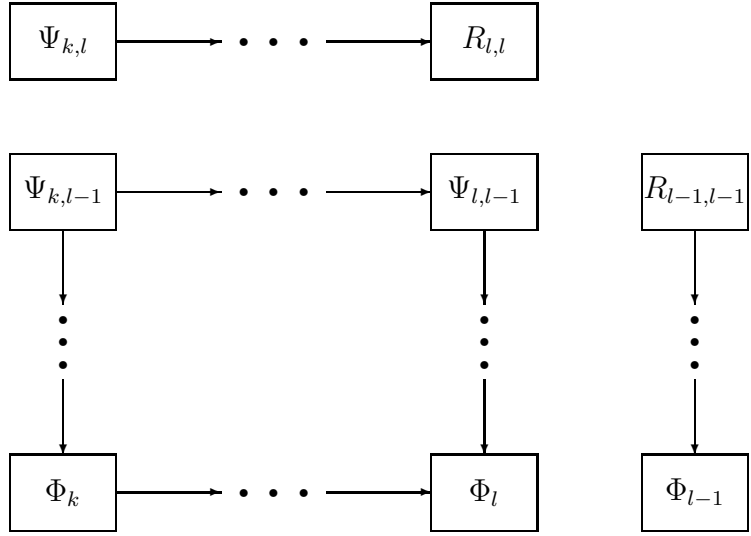
These fields describe a theory corresponding to irreducible
representation of de Sitter group with minimal number of degrees of
freedom, so from the de Sitter group point of view it can be called
massless. But in a flat limit it decomposes into a number of massless
representations of Lorentz group, the pattern now corresponds to
cutting boxes from the first row of Young tableau until we end up with
diagram having the same number of boxes in both rows. The rest fields
give partially massless theory for $Y(k+1,l)$ tensor. The reason is
that for description of massive theory we must have one more column of
fields as also shown on Figure 10. Note however that in this case we
obtain $d_{k,n}^2 < 0$ for all $n < l$ so this theory is non-unitary.

Besides these two cases we find a number of partially massless limits
which arise each time when one of the parameters $d_{k,n}$ (and hence
all parameters $d_{m,n}$) becomes equals to zero. At this point the
whole system also decomposes into two disconnected subsystems (and
diagram splits vertically into two blocks) as shown on Figure 11.
\begin{figure}[htb]
\begin{center}
\begin{picture}(132,90)
\put(1,1){\framebox(14,10)[]{$\Phi_k$}}
\put(8,17){\vector(0,-1){6}}
\multiput(8,19)(0,3){3}{\circle*{1}}
\put(8,33){\vector(0,-1){6}}
\put(1,33){\framebox(14,10)[]{$\Psi_{k,n}$}}
\put(15,6){\vector(1,0){10}}
\multiput(28,6)(4,0){3}{\circle*{1}}
\put(39,6){\vector(1,0){10}}
\put(49,1){\framebox(14,10)[]{$\Phi_l$}}
\put(15,38){\vector(1,0){10}}
\multiput(28,38)(4,0){3}{\circle*{1}}
\put(39,38){\vector(1,0){10}}
\put(49,33){\framebox(14,10)[]{$\Psi_{l,n}$}}
\put(56,17){\vector(0,-1){6}}
\multiput(56,19)(0,3){3}{\circle*{1}}
\put(56,33){\vector(0,-1){6}}

\put(69,1){\framebox(14,10)[]{$\Phi_{l-1}$}}
\put(76,17){\vector(0,-1){6}}
\multiput(76,19)(0,3){3}{\circle*{1}}
\put(76,33){\vector(0,-1){6}}
\put(69,33){\framebox(14,10)[]{$\Psi_{l-1,n}$}}
\put(83,6){\vector(1,0){10}}
\multiput(96,6)(4,0){3}{\circle*{1}}
\put(107,6){\vector(1,0){10}}
\put(117,1){\framebox(14,10)[]{$\Phi_{n}$}}
\put(83,38){\vector(1,0){10}}
\multiput(96,38)(4,0){3}{\circle*{1}}
\put(107,38){\vector(1,0){10}}
\put(117,33){\framebox(14,10)[]{$R_{n,n}$}}
\put(124,17){\vector(0,-1){6}}
\multiput(124,19)(0,3){3}{\circle*{1}}
\put(124,33){\vector(0,-1){6}}

\put(1,47){\framebox(14,10)[]{$\Psi_{k,n+1}$}}
\put(8,63){\vector(0,-1){6}}
\multiput(8,65)(0,3){3}{\circle*{1}}
\put(8,79){\vector(0,-1){6}}
\put(1,79){\framebox(14,10)[]{$\Psi_{k,l}$}}
\put(15,52){\vector(1,0){10}}
\multiput(28,52)(4,0){3}{\circle*{1}}
\put(39,52){\vector(1,0){10}}
\put(49,47){\framebox(14,10)[]{$\Psi_{l,n+1}$}}
\put(15,84){\vector(1,0){10}}
\multiput(28,84)(4,0){3}{\circle*{1}}
\put(39,84){\vector(1,0){10}}
\put(49,79){\framebox(14,10)[]{$R_{l,l}$}}
\put(56,63){\vector(0,-1){6}}
\multiput(56,65)(0,3){3}{\circle*{1}}
\put(56,79){\vector(0,-1){6}}

\end{picture}
\end{center}
\caption{Example of non-unitary partially massless limit in $dS$
space}
\end{figure}
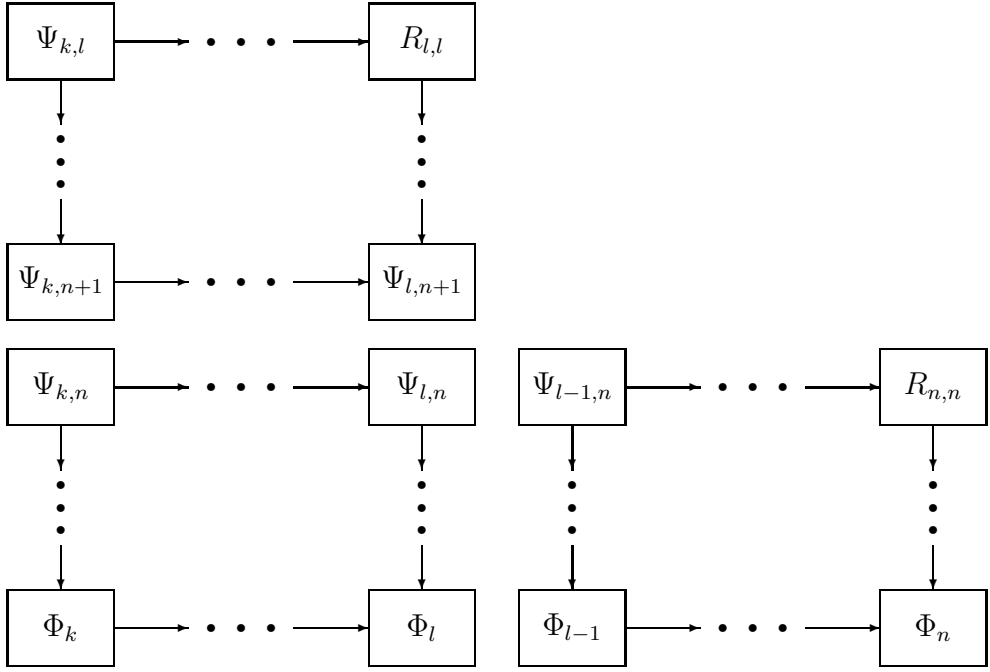
 The upper block gives one more example of partially massless theory
for the $Y(k+1,l+1)$ tensor, while the lower block describes
non-unitary (because $d_{k,p}^2 < 0$ for all $p < n$) partially
massless theory for the $Y(k+1,n)$ tensor we are already familiar
with. Indeed, to describe the massive theory for this tensor we need
one more block of fields also shown on Figure 11. In this, the same
partially massless theory can be obtained as a limit for massive
$Y(k+1,l+1)$ theory as well as a limit for massive $Y(k+1,n)$ theory
showing a consistency for the whole construction.

\section*{Conclusion}

Thus we have complete frame-like gauge invariant formulation both for
bosonic mixed symmetry tensors as well as fermionic mixed symmetry
spin-tensors corresponding to arbitrary Young tableau with two rows.
In all cases considered the same set of fields allows one to describe
massive theory in Minkowski space and (anti-)de Sitter spaces with
arbitrary cosmological constant. It is worth to note a great
similarity of results for bosonic and fermionic cases \cite{Zin09a}.
The main difference (the same as for completely symmetric
(spin-)tensors \cite{Zin08b}) is that in bosonic case we find first
partially limits at the boundaries of unitary forbidden regions while
for fermions all partially massless theories "live" inside the
forbidden regions.

\vskip 1cm
{\bf Acknowledgment} \\
Author is grateful to E. D. Skvortsov for stimulating discussions and
correspondence.

\appendix

\section{Explicit expressions for some complicated formulas}

Complete form of Formula (\ref{ap1}) in Section 1:
\begin{eqnarray*}
\delta_2 \Omega_{\mu\nu}{}^{(a_k),(b_l),c}&=& - 
\frac{2 a_{k,l}}{(k+2)(l+1)(d-4)}
 [ (k+1) l e_{[\mu}{}^c \xi_{\nu]}{}^{(a_k),(b_l)} -
 l e_{[\mu}{}^{(a_1} \xi_{\nu]}{}^{a_{k-1})c,(b_l)} + \nonumber \\
&&  \qquad \qquad \qquad +
e_{[\mu}{}^{(a_1} \xi_{\nu]}{}^{a_{k-1})(b_1,b_{l-1})c}
- (k+1) e_{[\mu}{}^{(b_1} \xi_{\nu]}{}^{(a_k),b_{l-1})c} + \\
 && \qquad \qquad \qquad + 
\rho_1 g^{(a_1a_2} \xi_{[\mu,\nu]}{}^{a_{k-2})(b_1,b_{l-1})c} +
\rho_2 g^{(a_1a_2} \xi_{[\mu,\nu]}{}^{a_{k-2})c,(b_l)} + \\
 && \qquad \qquad \qquad +
\rho_3 g^{(a_1a_2} \xi_{[\mu}{}^{a_{k-2})(b_1b_2,b_{l-2})c}{}_{\nu]} +
\rho_4 g^{(a_1a_2} \xi_{[\mu}{}^{a_{k-2})c(b_1,b_{l-1})}{}_{\nu]} + \\
 && \qquad \qquad \qquad +
\rho_5 g^{(a_1(b_1} \xi_{[\mu,\nu]}{}^{a_{k-1}),b_{l-1})c} +
\rho_6 g^{(a_1(b_1} \xi_{[\mu}{}^{a_{k-1})b_2,b_{l-2})c}{}_{\nu]} + \\
 && \qquad \qquad \qquad +
\rho_7 g^{(a_1(b_1} \xi_{[\mu}{}^{a_{k-1})c,b_{l-1})}{}_{\nu]} +
\rho_8 g^{(b_1b_2} \xi_{[\mu}{}^{(a_k),b_{l-2})c}{}_{\nu]} + \\
 && \qquad \qquad \qquad +
\rho_9 g^{c(b_1} \xi_{[\mu}{}^{(a_k),b_{l-1})}{}_{\nu]} +
\rho_{10} g^{c(a_1} \xi_{[\mu,\nu]}{}^{a_{k-1}),(b_l)} + \\
 && \qquad \qquad \qquad +
\rho_{11} g^{c(a_1} \xi_{[\mu}{}^{a_{k-1})(b_1,b_{l-1})}{}_{\nu]} ]
\end{eqnarray*}
$$
\rho_1 = \frac{2}{(d+k-4)}, \qquad
\rho_2 = - \frac{2l}{(d+k-4)}, \qquad
\rho_3 = - \frac{4}{(d+k-4)(d+l-5)}
$$
$$
\rho_4 = \frac{2(l-1)}{(d+k-4)(d+l-5)}, \qquad
\rho_5 = - \frac{k}{(d+k-4)}, \qquad
\rho_6 = \frac{(d+2k-4)}{(d+k-4)(d+l-5)}
$$
$$
\rho_7 = - \frac{[l(d+k-4) - k]}{(d+k-4)(d+l-5)}, \qquad
\rho_8 = - \frac{2(k+1)}{(d+l-5)}, \qquad
\rho_9 = \frac{[kl - (k-l+1)]}{(d+l-5)}
$$
$$
\rho_{10} = \frac{kl}{(d+k-4)}, \qquad
\rho_{11} = \frac{[(d+k-4) - kl]}{(d+k-4)(d+l-5)}
$$

$Y(k+1,n+1) \Leftrightarrow Y(k+1,n)$. Complete form for Formula
(\ref{ap2}) in Section 2.

\begin{eqnarray*}
\delta_1 \Psi_{\mu\nu}{}^{(a_k),(b_n)} &=& - 
\frac{d_{k,n}}{(k-n+2)(d+n-5)} \\
 && [ 
(k-n+1) \xi_{[\mu}{}^{(a_k),(b_{n-1}} e_{\nu]}{}^{b_1)} -
e_{[\nu}{}^{(a_1} \xi_{\mu]}{}^{a_{k-1})(b_1,b_{n-1})} + \\
 && +
\rho_1 g^{(a_1a_2} \xi_{[\mu,\nu]}{}^{a_{k-2})(b_1,b_{n-1})} +
\rho_4 g^{(a_1a_2} \xi_{[\mu}{}^{a_{k-2})(b_1b_2,b_{n-2})}{}_{\nu]} +
\\
 && + \rho_2 g^{(a_1(b_1} \xi_{[\mu,\nu]}{}^{a_{k-1}),b_{nl-1})} +
\rho_3 g^{(a_1(b_1} \xi_{[\mu}{}^{a_{k-1})b_2,b_{n-2})}{}_{\nu]} + \\
 && + \rho_5 g^{(b_1b_2} \xi_{[\mu}{}^{(a_k),b_{n-2})}{}_{\nu]}
\end{eqnarray*}
$$
\rho_1 = \frac{2}{(d+k+n-4)}, \qquad
\rho_2 = - \frac{(k-n)}{(d+k+n-4)}, \qquad
\rho_3 = \frac{(d+2k-4)}{(d+2n-6)(d+k+n-4)}
$$
$$
\rho_4 = - \frac{4}{(d+2n-6)(d+k+n-4)}, \qquad
\rho_5 = - \frac{2(k-n+1)}{(d+2n-6)}
$$
\begin{eqnarray*}
\delta_1 \Omega_{\mu\nu}{}^{(a_k),(b_n),c} &=& 
\frac{d_{k,n}}{(k-n+2)(d+n-4)} \\
 &&  \left[ (k-n+1) e_{[\mu}{}^{(b_1} \eta_{\nu]}{}^{(a_k),b_{n-1}),c}
-  e_{[\mu}{}^{(a_1} \eta_{\nu]}{}^{a_{k-1})(b_1,b_{n-1}),c} + \right.
\\
 && +
\rho_1 g^{(a_1a_2} \eta_{[\mu,\nu]}{}^{a_{k-2})(b_1,b_{n-1}),c} +
\rho_2 g^{(a_1a_2} 
\eta_{[\mu}{}^{a_{k-2})(b_1b_2,b_{n-2})}{}_{\nu]}{}^c + \\
 && + 
\rho_3 g^{(a_1a_2} 
\eta_{[\mu}{}^{a_{k-2})c(b_1,b_{n-1})}{}_{\nu]} + \rho_4 g^{(a_1a_2}
\eta_{[\mu}{}^{a_{k-2})(b_1b_2,b_{n-2})c}{}_{\nu]} + \\
 && +
\rho_5 g^{(a_1(b_1} \eta_{[\mu,\nu]}{}^{a_{k-1}),b_{n-1}),c} +
\rho_6 g^{(a_1(b_1} \eta_{[\mu}{}^{a_{k-1})b_2,b_{n-2})}{}_{\nu]}{}^c
+ \\
 && + \rho_7 g^{(a_1(b_1} 
\eta_{[\mu}{}^{a_{k-1})b_2,b_{n-2})c}{}_{\nu]} + \rho_8 g^{(a_1(b_1}
\eta_{[\mu}{}^{a_{k-1})c,b_{n-1})}{}_{\nu]} + \\
 && +
\rho_9 g^{c(a_1} \eta_{[\mu}{}^{a_{k-1})(b_1,b_{n-1})}{}_{\nu]}
+ \rho_{10} g^{c(b_1} \eta_{[\mu}{}^{(a_k),b_{n-1})}{}_{\nu]} + \\
 &&  + \left.
\rho_{11} g^{(b_1b_2} \eta_{[\mu}{}^{(a_k),b_{n-2})}{}_{\nu]}{}^c
+ \rho_{12} g^{(b_1b_2} \eta_{[\mu}{}^{(a_k),b_{n-2})c}{}_{\nu]}
\right]
\end{eqnarray*}
$$
\rho_1 = - \frac{2}{(d+k+n-4)}, \qquad
\rho_5 = \frac{(k-n)}{(d+k+n-4)}, \qquad
\rho_9 = - \frac{1}{(d+n-5)}
$$
$$
\rho_2 = \frac{4}{(d+2n-6)(d+k+n-4)}, \qquad
\rho_3 = \frac{2}{(d+n-5)(d+k+n-4)}
$$
$$
\rho_4 = -  \frac{4}{(d+n-5)(d+2n-6)(d+k+n-4)}, \qquad
\rho_6 = - \frac{(d+2k-4)}{(d+2n-6)(d+k+n-4)}
$$
$$
\rho_7 = \frac{(d+2k-4)}{(d+n-5)(d+2n-6)(d+k+n-4)}, \qquad
\rho_8 = - \frac{(k-n)}{(d+n-5)(d+k+n-4)},
$$
$$
\rho_{10} = \frac{(k-n+1)}{(d+n-5)}, \qquad
\rho_{11} = \frac{2(k-n+1)}{(d+2n-6)}, \qquad
\rho_{12} = - \frac{2(k+n-1)}{(d+n-5)(d+2n-6)}
$$

$Y(m+1,n+1) \Leftrightarrow Y(m,n+1)$. Complete form for Formula
(\ref{ap3}) in Section 3.

\begin{eqnarray*}
\delta_1 \Psi_{\alpha\beta}{}^{(a_m),(b_n)} &=&
 - \frac{c_{m,n}}{(d+m-4)} [
e_{[\alpha}{}^{(a_1} \xi_{\beta]}{}^{a_{m-1}),(b_n)} + 
\rho_1 g^{(a_1a_2} \xi_{[\alpha,\beta]}{}^{a_{m-2}),(b_n)} + \\
 && \qquad \qquad \qquad + \rho_2 g^{(a_1a_2} 
\xi_{[\alpha}{}^{a_{m-2})(b_1,b_{n-1})}{}_{\beta]} + \rho_3
g^{(a_1(b_1} \xi_{[\alpha}{}^{a_{m-1}),b_{n-1})}{}_{\beta]} ]
\end{eqnarray*}
$$
\rho_1 = \frac{2}{(d+2m-4)}, \qquad 
\rho_2 = - \frac{2}{(d+2m-4)(d+m+n-4)}, \qquad
\rho_3 = \frac{1}{(d+m+n-4)}
$$
\begin{eqnarray*}
\delta_1 \Omega_{\mu\nu}{}^{(a_m),(b_n),c} &=& - 
\frac{c_{m,n}}{(d+m-3)} [
 e_{[\mu}{}^{(a_1} \eta_{\nu]}{}^{a_{m-1}),(b_n),c} +
\rho_1 g^{(a_1a_2} \eta_{\mu,\nu}{}^{a_{m-2}),(b_n),c} + \\
 && \qquad \qquad \qquad + \rho_2 g^{(a_1a_2} 
\eta_{[\mu}{}^{a_{m-2})(b_1,b_{n-1})}{}_{\nu]}{}^c + \rho_3
g^{(a_1a_2} \eta_{[\mu}{}^{a_{m-2})(b_1,b_{n-1})c}{}_{\nu]} + \\
 && \qquad \qquad \qquad + 
\rho_7 g^{(a_1a_2} \eta_{[\mu}{}^{a_{m-2})c,(b_n)}{}_{\nu]}  +
\rho_4 g^{(a_1(b_1} \eta_{[\mu}{}^{a_{m-1}),b_{n-1})}{}_{\nu]}{}^c +
\\
 && \qquad \qquad \qquad + 
\rho_5 g^{(a_1(b_1} \eta_{[\mu}{}^{a_{m-1}),b_{n-1})c}{}_{\nu]}
 + \rho_6 g^{c(a_1} \eta_{\mu}{}^{a_{m-1}),(b_n)}{}_{\nu]} ] 
\end{eqnarray*}
$$
\rho_1 = \frac{2}{(d+2m-4)}, \qquad 
\rho_2 = - \frac{2}{(d+2m-4)(d+m+n-4)}, \qquad
\rho_4 = \frac{1}{(d+m+n-4)}
$$
$$
\rho_3 = \frac{2}{(d+m-4)(d+2m-4)(d+m+n-4)}, \qquad
\rho_7 = - \frac{2}{(d+m-4)(d+2m-4)}
$$
$$
\rho_5 = - \frac{1}{(d+m-4)(d+m+n-4)}, \qquad
\rho_6 = \frac{1}{(d+m-4)}
$$

\end{document}